\documentclass[12pt,preprint]{aastex}

\slugcomment{Submitted to the Astrophysical Journal}
\shorttitle{Multiple Weak Deflections}
\shortauthors{Brainerd}

\begin{document}

\title{Multiple Weak Deflections in Galaxy-Galaxy Lensing}

\author{Tereasa G.\ Brainerd}
\affil{Boston University, Institute for Astrophysical Research, 725
Commonwealth Ave., Boston, MA 02215}
\email{brainerd@bu.edu}

\begin{abstract}
The frequency and effects of multiple weak deflections
in galaxy-galaxy lensing are investigated via Monte Carlo simulations.
The lenses in the simulations are galaxies with known
redshifts and known rest-frame blue luminosities.
The frequency of multiple
deflections above a given threshold shear value is quantified for 
discrete source redshifts, as well as for a set of sources that
are broadly distributed in redshift space.  In general, the
closest lens in projection on the sky is not the only lens for
a given source. In addition, $\sim 50$\% of the time the closest
lens is not the most important lens for
a given source.  Compared to a naive single-deflection
calculation in which only the lensing due to the closest weak lens is considered, 
a full multiple-deflection
calculation yields a higher net shear for individual sources, as well as 
a higher mean tangential shear around the lens centers.  The
full multiple-deflection calculation also
shows that galaxy-galaxy lensing may contribute
a substantial amount to cosmic shear on small angular scales.  The
degree to which galaxy-galaxy lensing contributes to the small-scale
cosmic shear is, however, quite sensitive
to the mass adopted for the halos of $L_B^\ast$ galaxies.  Changing
the halo mass by a factor of $\sim 2.5$ changes the contribution of galaxy-galaxy lensing
to the cosmic shear by a factor of $\sim 3$ on scales of 
$\theta \sim 1$~arcmin.  The contribution of galaxy-galaxy lensing to cosmic
shear decreases rapidly with angular scale and extrapolates to zero
at $\theta \sim 5$~arcmin.  This last result is roughly independent of the
halo mass and suggests that for scales $\theta \gtrsim 5$~arcmin, cosmic
shear is insensitive to the details of the gravitational potentials 
of large galaxies.
\end{abstract}


\section{Introduction}

Galaxy-galaxy lensing is the systematic weak gravitational lensing of background
galaxies by foreground galaxies.  Unlike weak lensing by massive galaxy clusters,
where the only important lens in the problem is the cluster itself,
galaxy-galaxy lensing involves multiple weak deflections. 
For example, it is common for
a distant source galaxy at redshift $z_s$ to be weakly lensed by a more
nearby galaxy at redshift
$z_{l1}$, and for both of these galaxies to then be lensed by another (even more 
nearby) galaxy at redshift $z_{l2}$.  Thus, the galaxy with redshift
$z_{l1}$ serves simultaneously as a lens for the galaxy at $z_s$ and
a source for the galaxy at $z_{l2}$.  In addition, the galaxy at $z_s$ is
lensed by two independent foreground galaxies.  The importance of such multiple
deflections in galaxy-galaxy lensing was first noted by
Brainerd et al.\ (1996; hereafter BBS).
Since the work of BBS, galaxy-galaxy lensing has been detected
with impressively high statistical significance by a number of 
different groups.  This has enabled
constraints to be placed on the nature of the dark
matter halos that surround the lens galaxies as well as the bias between mass and light
in the universe (see, e.g., 
Fischer et al.\ 2000; Guzik \& Seljak 2002; Hoekstra et al.\ 2004, 2005; Sheldon
et al.\ 2004; Heymans et al.\ 2006; Kleinheinrich et al.\ 2006; Mandelbaum et al.\
2006ab; Limousin et al.\ 2007; Parker et al.\ 2007; Natarajan et al.\ 2009; Tian et al.\ 2009
).

The purpose of the present investigation is to: (i) quantify the frequency
of multiple weak lensing deflections in a relatively deep galaxy-galaxy
lensing data set, (ii) 
determine the effect of multiple deflections on the net shear for distant
source galaxies that have been weakly lensed by 
foreground galaxies, and (iii) demonstrate that galaxy-galaxy lensing alone 
may contribute a substantial amount to the ``cosmic shear'' signal on small
angular scales.  To do this,
theoretical shear fields are constructed using a set of observed galaxies 
with known redshifts and known rest-frame blue luminosities.  A simple
halo model is used to assign masses to the observed galaxies and Monte
Carlo simulations are then used to lens various 
theoretical source galaxy distributions by the observed galaxies.  Theoretical
shear fields for full, multiple-deflection calculations are computed;
i.e., each source
galaxy in the simulation is lensed by all foreground galaxies.  In addition, 
theoretical shear fields for naive, single-deflection calculations (where
the closest lens on the sky is assumed to be the only lens) are also computed.
The results of the single-deflection calculations are compared to those of
the full, multiple-deflection calculations in order to assess the effects
of multiple deflections in galaxy-galaxy lensing.  Throughout, the weak
lensing of an entire source galaxy by a single foreground lens galaxy will
be referred to as a ``deflection''.

The paper is organized as follows.  The Monte Carlo simulations of
galaxy-galaxy lensing are described in Section 2, the frequency of multiple
weak deflections in galaxy-galaxy lensing is computed in Section 3, the 
effects of multiple weak deflections on the galaxy-galaxy lensing
shear are computed in Section 4, the contribution of galaxy-galaxy lensing
to cosmic shear is computed in Section 5, and a discussion of cosmic
variance in relation to the field size in presented in Section 6.  
The conclusions are summarized in
Section 7.

\section{Monte Carlo Simulations of Galaxy-Galaxy Lensing}

To investigate the frequency and effects of multiple weak deflections
in galaxy-galaxy lensing, Monte Carlo simulations are constructed.
The lens galaxies in the Monte Carlo simulations are 
relatively bright galaxies that are contained within a circle
of radius 4~arcmin, centered on the Hubble Deep Field North (HDF-N) 
(Williams et al.\ 1996).  This region of sky was the subject
of a deep redshift survey (Cohen et al.\ 1996; Steidel et al.\ 1996;
Lowenthal et al.\ 1997; Phillips et al.\ 1997; Cohen et al.\ 2000)
as well as an extensive multicolor
photometric investigation (Hogg et al.\ 2000).  As a result, both the redshifts
(Cohen et al.\ 2000, Tables~2A and 2B)
and rest-frame blue luminosities, $L_B$ (Cohen 2001, Table~1), of 590 galaxies in this 
region of the sky are known.  
For the simulations, then, the locations of the lenses in redshift
space are known very accurately, and the relative strengths of the different lenses
can be inferred quite well from their relative luminosities.
Therefore, it is possible to make
detailed theoretical predictions for the weak galaxy-galaxy lensing shear field
that should be expected in this region of sky.

For simplicity, the dark matter halos of the lens galaxies are taken to have
a mass density given by
\begin{equation}
\rho(r) = \frac{\sigma_v^2 s^2}{2\pi G r^2 \left(r^2 + s^2 \right)},
\end{equation}
where $\sigma_v$ is the velocity dispersion of the halo, $G$ is Newton's constant,
and $s$ is a characteristic halo radius (see e.g., BBS; Hudson et al.\ 1998;
Fischer et al.\ 2000; Hoekstra et al.\ 2004).  It is then
convenient to scale the depths of the potential wells of lens galaxies with differing
luminosities, $L_B$, according to a Tully-Fisher or Faber-Jackson type of
relation
\begin{equation}
\frac{\sigma_v}{\sigma_v^\ast} = \left( \frac{L_B}{L_B^\ast} \right)^{1/4},
\end{equation}
where $\sigma_v^\ast$ is the velocity dispersion of the halo of a lens galaxy 
with rest-frame blue luminosity $L_B^\ast$.  Again for simplicity, it is 
assumed that the mass-to-light ratio of a galaxy is constant independent of
its luminosity. Therefore, the characteristic radii of the halos of galaxies with
$L_B \ne L_B^\ast$ scale with the radii of the halos of $L_B^\ast$ galaxies as
\begin{equation}
\frac{s}{s^\ast} = \left( \frac{L_B}{L_B^\ast} \right)^{1/2} .
\end{equation}
Under these assumptions, then, the mass of the halo of an $L_B^\ast$ galaxy is
given by
\begin{equation}
M^\ast = \frac{\pi s^\ast ( \sigma_v^\ast )^2}{G} 
\end{equation}
and 
the deflection of a light ray emanating from a source galaxy is given by
\begin{equation}
\alpha(X) = \frac{4\pi \sigma_v^2 D_{ls}}{D_s X c^2} \left[ 1 + X -
\left( 1 + X^2 \right)^{1/2} \right].
\end{equation}
Here $D_s$ is the angular diameter distance between the
observer and the source, $D_{ls}$ is the angular diameter distance between the 
lens and the source, and $X$ is the ratio of the impact parameter of the
light ray and the characteristic radius, $s$, of the lens
(i.e., $X \equiv R/s$; see BBS).   

It is worth noting that galaxy-galaxy lensing has, of course, been detected in the 
HDF-N (e.g., dell'Antonio \& Tyson 1996; Hudson et al.\ 1998); however due to the very small
number of galaxies in the HDF-N, the galaxy-galaxy lensing signal can only be
detected with relatively low significance.  In particular, there are simply too
few actual source galaxies 
to carry out a detailed investigation
of the effects of multiple deflections using only the
observed sources.  It is for this reason that Monte Carlo
simulations are adopted here. 

The completeness limits of the redshift survey are, unfortunately, different for
the HDF-N itself and the surrounding area of the sky,
the survey being deeper in the region of the HDF-N.
This gives rise to a somewhat different redshift distribution
for galaxies with measured redshifts in the center of the field versus galaxies
with measured redshifts in the outer
region of the field.
In order to make an accurate prediction for the theoretical shear field, it is important
that the redshift completeness limit for the lenses in the Monte Carlo simulations be uniform
across the field.
Therefore, a conservative
completeness limit of $R = 23$ is imposed here, and the lenses in the Monte Carlo simulations
consist of the 427 galaxies with $R \le 23$ in Cohen et al.\ (2000) and Cohen (2001)
for which spectroscopic redshifts and rest-frame blue luminosities are known.  The median
redshift of the lens galaxies is therefore $z_{\rm med} = 0.55$.

Two approaches are taken to model the redshifts of
the source galaxy population: (i) source galaxies are
simply placed in a single plane of redshift $z_s$ and (ii) source galaxies are distributed
in redshift space according to the observed redshift distribution of faint galaxies.  The
first approach allows an investigation of the frequency of multiple weak deflections as 
a function of discrete source redshift.  The second approach demonstrates the overall 
effect that would be expected to occur in a deep galaxy-galaxy lensing data set.

Each Monte Carlo simulation includes
10 million source galaxies that are assigned random positions (RA and DEC) within a
circle of radius 2.5~arcminutes, centered on the HDF-N.   The sources are contained within
a smaller area than the lenses because, as will be shown below, about 10\% of the time
the most important lens for a given source may be more than an arcminute away.
By restricting the sources to a smaller area than the lenses, edge effects
(in which sources are not properly lensed by all foreground galaxies) are avoided.
In simulations where the sources are restricted
to a single plane in redshift space, 
each source is assigned the identical redshift, $z_s$.  In simulations
where the sources are broadly distributed in redshift space, the apparent magnitudes of the sources
are taken to be in the range $19 < I < 25$, and the number of sources per unit magnitude
is chosen to match the observed number counts in the $I$-band (e.g., Smail et al.\ 1995).  These
sources are assumed to follow a redshift distribution of the form
\begin{equation}
P(z|I) = \frac{\beta z^2 \exp \left[ - (z/z_0)^\beta \right]}{\Gamma(3/\beta) z_0^3},
\end{equation}
which is in good agreement with the redshift surveys of LeF\`evre et al.\ (1996) and
LeF\`evre et al.\ (2004).  Assuming $\beta = 1.5$ and
extrapolating the results of LeF\`evre et al.\ (2004) to a sample of galaxies with
$19 < I < 25$ yields
\begin{equation}
z_0 = 0.8 \left[0.86 + 0.15(I-23.35) \right]
\end{equation}
(see, e.g., BBS).  The median redshift of the sources in this case is $z_{\rm med} = 0.96$.

Throughout, we will consider only the weak lensing
regime. That is, we will restrict our analysis to the case that
the surface mass density of the lenses is very much less
than the critical surface mass density for strong lensing
($\Sigma(\theta) << \Sigma_c \equiv \frac{c^2}{4\pi G} \frac{D_s}{D_d D_{ls}}$),
the deflection angle, $\alpha$, the modulus of the
shear, $\gamma$, and the convergence,
$\kappa$, are all small, and $\gamma \simeq \kappa$.  Given that the physical size of 
each lens is very much smaller than the distances between the observer, lens, and
source, we will adopt the standard thin lens approximation (e.g., Blandford \& Narayan
1986; Schneider et al.\ 1992).
Further, we will perform all calculations in the 
framework of the Born approximation, in which integrations are performed along an
undeflected light ray.  This standard weak lensing formalism is valid even 
in the limit of multiple weak deflections (e.g., Bartelmann \& Schneider 2001).  Indeed,
investigations into the degree to which the Born approximation may affect predictions
of cosmic shear (where the weak lenses consist of all the mass along the line of sight),
have shown that corrections due to the Born approximation are two to three orders
of magnitude smaller than the cosmic shear signal itself (e.g., Cooray \& Hu 2001;
Hilbert et al.\ 2009).

For each Monte Carlo simulation, specific values of
the velocity dispersion, $\sigma_v^\ast$, and characteristic
radius, $s^\ast$, for $L_B^\ast$ galaxies are chosen.  Velocity dispersions, $\sigma_v$, and
characteristic radii, $s$, are then assigned to each lens galaxy based upon the above scaling relations.
The redshifts of the lenses, $z_l$, are taken to be
the observed spectroscopic redshifts, and the positions
of the lenses in the field (RA and DEC) are taken to be
the observed positions on the sky.  The Monte Carlo simulation
then proceeds by computing the weak lensing shear, $\vec{\gamma}$, that
is induced as the light rays emanating from
the background sources encounter the foreground lenses.  
In the case of 
single-deflection calculations, the lensing of each source is 
computed solely for the lens which 
is nearest to the source in projection on the sky.  That is, the ``closest'' lenses are the only lenses
that are used in the single deflection calculations, and the resulting shear for each source is simply the
shear induced by the closest lens.
In the case of full, multiple-deflection
calculations, the lensing of each source by all foreground lenses is computed.   
The resulting shear for each source is then the net shear due to all foreground 
lenses.  This is straightforward to compute 
in the weak lensing regime since all weak deflections
may be considered to be independent (e.g., Bartelmann \& Schneider 2001).

Each source galaxy is assigned a random intrinsic position angle and
an intrinsic ellipticity that is drawn at random from the observed ellipticity distribution
of the HDF-N galaxies.  The intrinsic shape parameters of the source galaxies are
then given by
\begin{equation}
\vec{\chi}_{in} = \epsilon_{in}~ e^{2i \phi_{in}}
\end{equation}
where $\epsilon_{in} = (a-b)/(a+b)$ is the intrinsic ellipticity of the source
and $\phi_{in}$ is its intrinsic position angle.  
Since we are dealing only with the weak lensing regime, the final
image shape of each source galaxy in the multiple-deflection calculations is given by
\begin{equation}
\vec{\chi}_f = \vec{\chi}_{in} + \Sigma_{j=1}^{N_{\rm lens}}  ~ \vec{\gamma}_j
\end{equation}
where $\vec{\gamma}_j$ is the shear induced by foreground lens galaxy, $j$.
In the case of the full, multiple-deflection calculations, the net shear due to 
all lenses with $z_l < z_s$ is used to obtain $\vec{\gamma}_f$ for each source galaxy.
In the case of the single deflection calculations, the sum over all foreground lenses is
simply replaced by $\vec{\gamma}_{\rm close}$, the shear induced by the lens that
is closest to the source in projection on the sky.

Shown in Figure 1 is a zoomed-in image of one of the simulations.  The image is
centered on the HDF-N, and the locations of
chips 2, 3, and 4 on WFPC-2 are shown by the black lines.  Here
a fiducial lens halo model with $\sigma_v^\ast = 150$~km~sec$^{-1}$ and $s = 100h^{-1}$~kpc 
has been adopted, and the source galaxies have been distributed in redshift space according to
equation (6) above.  A flat
$\Lambda$-dominated cosmology with $H_0 = 70$~km~sec$^{-1}$~Mpc$^{-1}$,
$\Omega_{m0} = 0.3$ and $\Omega_{\Lambda 0} = 0.7$ has been also been adopted.  
The left panel of Figure 1 shows the magnitude of the net shear, and for clarity
the orientation of the net shear is not shown.
Red peaks in the shear field
(i.e., locations of the largest net shear) correspond to the locations of the 
most important weak galaxy lenses in the field. 
The right panel of Figure 1 shows the surface mass density of the lens galaxies.
Note that some very luminous (and, therefore very massive) galaxies do not show up
in the shear field due to the fact that their redshifts place them well beyond the
median redshift of the sources.  A good example of this is the galaxy 
located at $(-23.20,+56.70)$ in Figure~1.  This galaxy has coordinates on the sky of
${\rm RA} =12^h~ 36^m~ 52.72^s, {\rm DEC} =+62^\circ~ 13'~ 54.70''$ (J2000).
Its rest-frame
blue luminosity is $2.95L_B^\ast$ and, hence, its halo
mass is $6.9\times 10^{12} M_\odot$ (for the fiducial model).  The center
of this galaxy has a high surface mass density (indicated by red in the
right panel of Figure~1). However, since the redshift of this galaxy
is $z = 1.355$, it cannot act as a lens for the majority of the sources.  Therefore, it does
not contribute substantially
to the net shear field.  By contrast, the two smaller galaxies that are immediately to
the east and west of this intrinsically very bright and massive galaxy 
do show up quite prominently in the shear field.
These galaxies have coordinates on the sky of
${\rm RA} =12^h~ 36^m~ 54.07^s, {\rm DEC}=+62^\circ~ 13'~ 54.20''$ 
and
${\rm RA} =12^h~ 36^m~ 51.77^s, {\rm DEC}=+62^\circ~ 13'~ 53.70''$,
corresponding to
locations of $(-32.65,+56.20)$ and $(-16.56,+55.70)$ in Figure 1.  These two
galaxies have luminosities of $L_{\rm east} = 0.70 L_B^\ast$ and
$L_{\rm west} = 0.87 L_B^\ast$, and
redshifts of $z_{\rm east} = 0.851$ and
$z_{\rm west} = 0.557$.  Both of these galaxies are
assigned very similar halo masses in the simulation
(since their luminosities are very similar),
and both are clearly visible in the shear field as red peaks.  However,
the easternmost of these two galaxies corresponds to a smaller peak in
the shear field than the westernmost because the redshift of the easternmost
galaxy is only slightly less than the median redshift of the sources,
while the redshift of the westernmost galaxy is of order half the median 
redshift of the sources.  For a color image of the HDF-N in which the redshifts
of the galaxies are indicated, the reader is encouraged to see Figure~2 of
Cohen et al.\ (2000).

\begin{figure}
\centerline{\scalebox{1.20}{\includegraphics{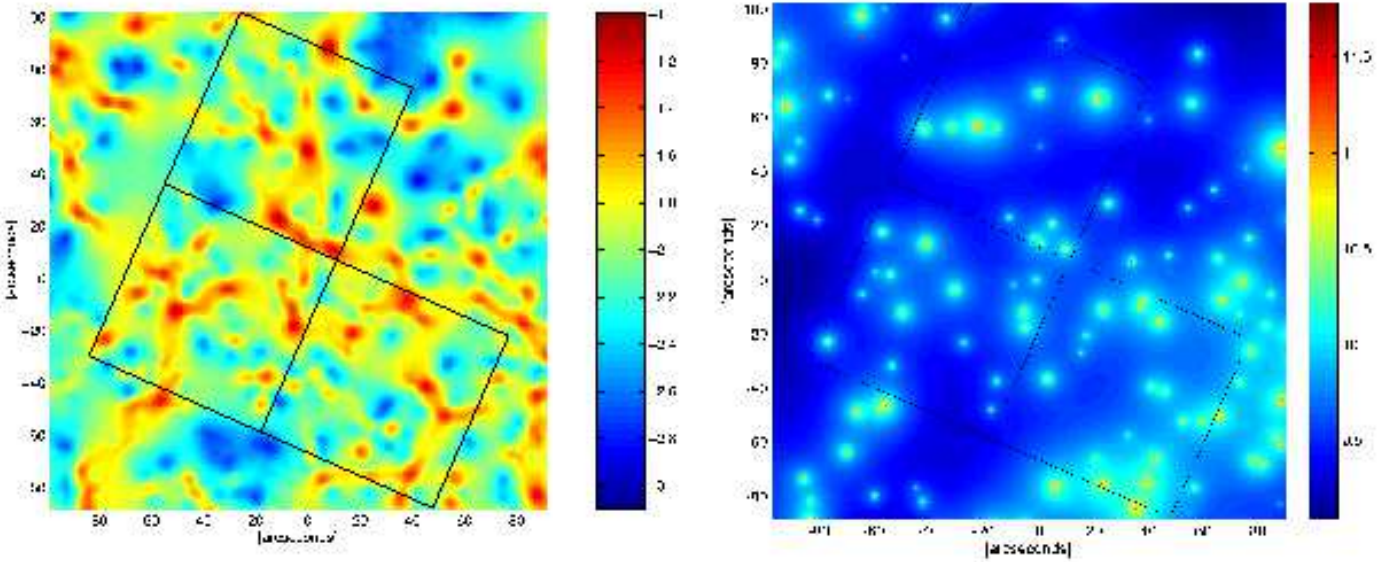} } }%
\vskip -0.0cm%
\caption{
A localized region of a simulation, centered on the HDF-N.  The figure
has been oriented according to the standard convention
(i.e., ``North'' is up, and ``East'' is to the left).
The characteristic chevron
of WFPC-2 is indicated by the black lines.
Note that in the full simulations, the lens galaxies
are contained within an area that is $\sim 16.5$ times larger than the
HDF-N. 
A fiducial halo model with 
$\sigma_v^\ast = 150$~km~sec$^{-1}$ and $s^\ast = 100h^{-1}$~kpc, and cosmological parameters
$H_0 = 70$~km~sec$^{-1}$~Mpc$^{-1}$, $\Omega_{m0}$, and $\Omega_{\Lambda 0}$ have been adopted.
The median redshift of the lenses is $z_l = 0.55$ and the median redshift of the
sources is $z_s = 0.96$.  
Left: Logarithm of the
net shear produced by the lens galaxies.  
Peaks in the shear field correspond to the most 
important weak galaxy lenses
in the localized region of the HDF-N.  Right:
Logarithm of the surface mass density 
of the lens galaxies.
Here the units of surface mass density
are solar masses per square arcsecond.
Due to their redshifts being much greater than
the median redshift of the sources,
some galaxies that contribute
significantly to the surface mass density do not contribute significantly
to the shear field.  Conversely, some galaxies that contribute 
relatively little to the surface mass density contribute a substantial amount
to the shear field because their redshifts are considerably 
smaller than the median redshift of the sources.
}
\label{fig1}
\end{figure}

\section{Frequency of Multiple Deflections}

The probability that a given source galaxy will have been
weakly-lensed by one or more foreground galaxies is, of course, a strong
function of the actual value of the shear, $\gamma$, induced by a given
weak lensing deflection.  That is, it is
much more likely for a distant galaxy to be lensed by a foreground
galaxy which
produces an insignificant weak shear of
$\gamma \sim 10^{-6}$ than, say, a relatively large weak
shear of $\gamma \sim 0.01$.  Therefore, in order to discuss the total number
of weak deflections that a given source galaxy is likely to encounter, a
decision has to be made as to what value of $\gamma$ qualifies as a
``significant'' value of the shear.
A typical value of the shear induced by a single weak galaxy lens is
$\gamma \sim 0.005$ (see, e.g., BBS)
and this value of $\gamma$ will be used as a baseline
for computing the number of weak lensing deflections that source galaxies
have undergone in the Monte Carlo simulations.  

To begin this section, the Monte Carlo 
simulations will be restricted to the fiducial halo lens model from the previous
section in which $\sigma_v^\ast = 150$~km~sec$^{-1}$ and $s^\ast = 100h^{-1}$~kpc,
and sources will be placed in single planes in redshift (i.e., all sources will
be assigned identical redshifts).
Figures 2, 3, and
4, then, show the probability, $P(N_L)$,
that a given source with redshift $z_s$
will be lensed by $N_L$ foreground galaxies,
where each {\it individual deflection} gives rise to
a shear of $\gamma > 0.0025$, $\gamma > 0.005$, and $\gamma > 0.01$, respectively
(i.e., the minimum shear in these figures
corresponds to half the baseline value, the baseline
value, and twice the baseline value, respectively).   Here
$P(N_D = 2)$ is the probability that
a given source will be lensed by two individual
foreground galaxies, each of which
lensed the source galaxy at a level that is comparable to or greater
than the minimum shear value.  Since the minimum values adopted in Figures
2, 3, and 4 are ``substantial'' values of the galaxy-galaxy lensing shear,
the results shown in these figures are conservative estimates
of the frequency of multiple deflections.  The line types in Figures 2, 3, 
and 4 correspond to different values of the cosmological parameters.  In 
all cases $H_0 = 70$~km~sec$^{-1}$~Mpc$^{-1}$ is adopted.  Solid lines
show results for a flat $\Lambda$-dominated universe with 
$\Omega_{m0} = 0.3$ and $\Omega_{\Lambda 0} = 0.7$, dashed lines show results for
an open universe with $\Omega_{m0} = 0.3$ and $\Omega_{\Lambda 0} = 0.0$, and
dotted lines show results for an Einstein-de Sitter universe.

\begin{figure}
\centerline{\scalebox{0.70}{\includegraphics{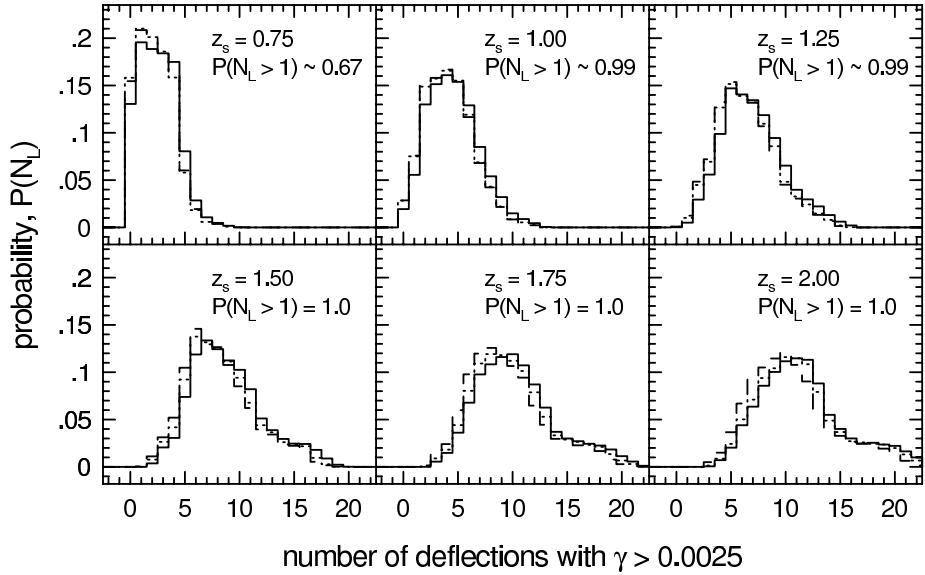} } }%
\vskip -0.0cm%
\caption{
Probability, $P(N_L)$, that a source galaxy 
with redshift $z_s$ will be lensed by
$N_L$ foreground galaxies, where each individual
lens induces a shear $\gamma > 0.0025$.
For $N_L > 1$, multiple deflections with $\gamma > 0.0025$ have been 
experienced by the source.  Source redshifts range from 
$z_s = 0.75$ (top left) to $z_s = 2.0$ (bottom right).  The median lens redshift
is $z_l = 0.55$.  A fiducial halo model with $\sigma_v^\ast = 150$~km~sec$^{-1}$
and $s^\ast = 100h^{-1}$~kpc has been adopted.  Line types correspond to different
values of the cosmological parameters.  Solid lines: flat $\Lambda$-dominated universe.
Dashed lines: open universe.  Dotted lines: Einstein-de Sitter universe.
}
\label{fig2}
\end{figure}

\begin{figure}
\centerline{\scalebox{0.70}{\includegraphics{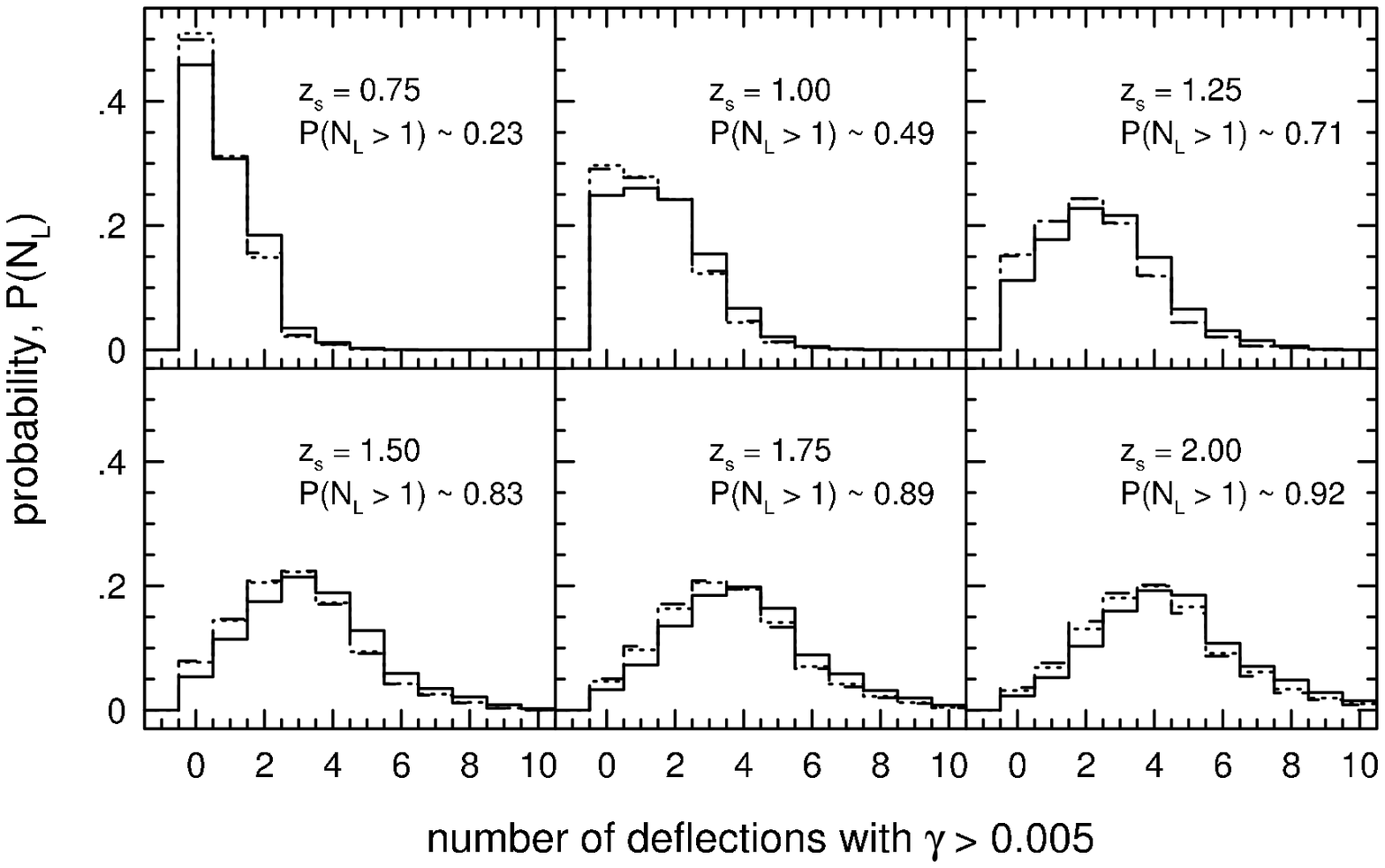} } }%
\vskip -0.0cm%
\caption{
Same as Figure 2, except here the frequency of deflections with
$\gamma > 0.005$ is shown.  For $N_L > 1$, multiple deflections
have been experienced by the source.
}
\label{fig3}
\end{figure}

\begin{figure}
\centerline{\scalebox{0.70}{\includegraphics{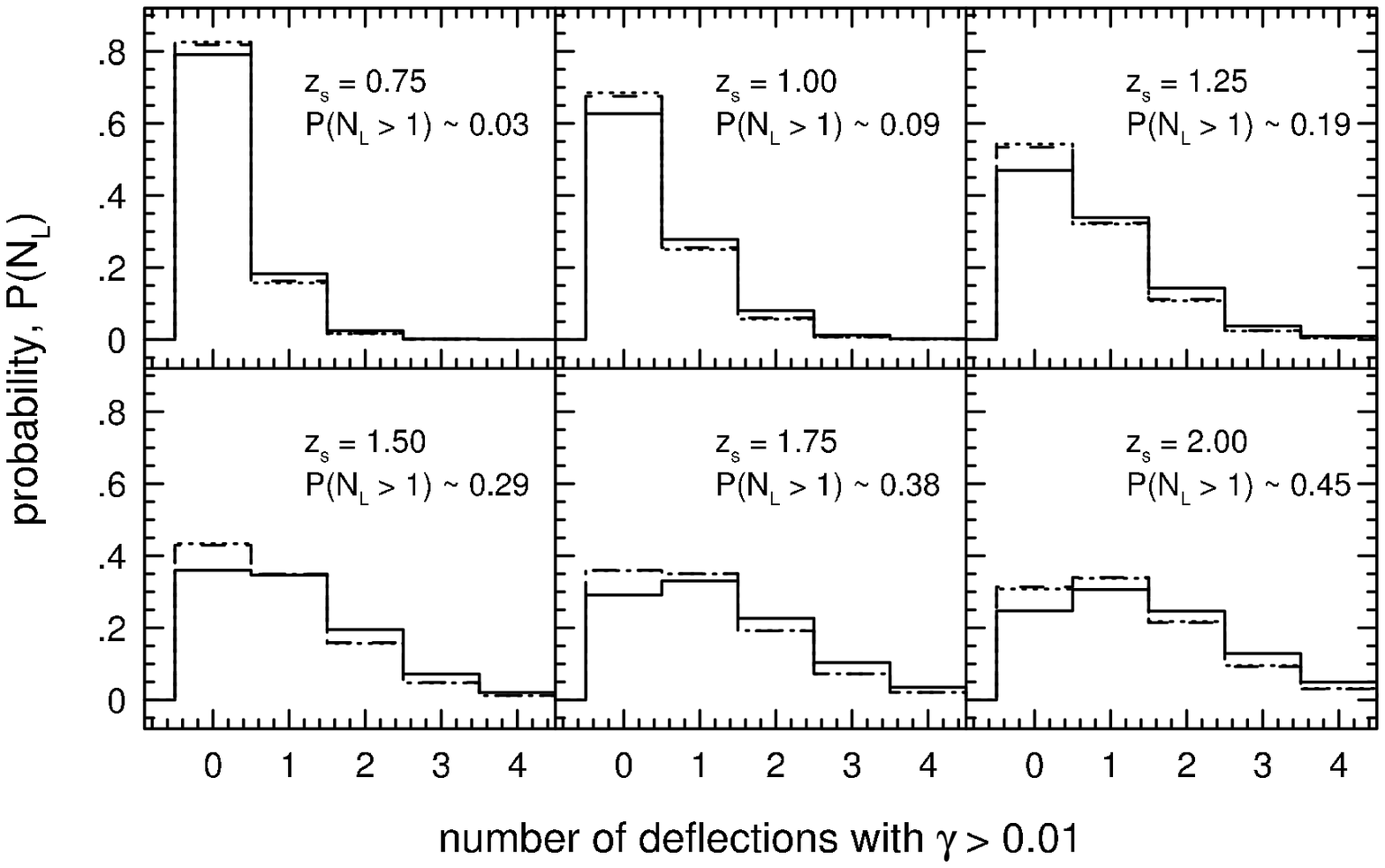} } }%
\vskip -0.0cm%
\caption{
Same as Figure 2, except here the frequency of deflections with
$\gamma > 0.01$ is shown.  For $N_L > 1$, multiple deflections
have been experienced by the source.
}
\label{fig4}
\end{figure}

Figures 2, 3, and 4 demonstrate two fully expected results.
First, the frequency of multiple deflections in
galaxy-galaxy lensing is
a function of the source redshift: the higher the redshift, the more likely
multiple deflections are to occur.   Second, 
the frequency of multiple deflections in galaxy-galaxy 
lensing depends upon the minimum shear value 
that is adopted: the lower the value of the minimum shear, the more likely
that multiple deflections of at least the minimum value will occur.  Figure 2
shows that multiple deflections in which each individual deflection results in
a shear of $\gamma > 0.0025$
are highly probable.  The probability ranges from 
67\% for sources with $z_s = 0.75$ to 100\% for sources with $z_s = 2.0$.  
Similarly, Figure 3 shows that multiple deflections in which each individual
deflection results in a shear of $\gamma > 0.005$ are highly probable.  In this
case, the probability ranges from 23\% for sources with $z_s = 0.75$ to
92\% for sources with $z_s = 2.0$.  Multiple deflections in which each individual
deflection results in a shear of $\gamma > 0.01$ are relatively rare for sources
with $z_s \le 1.0$, but the probability of such very large multiple deflections
increases to 45\% for sources with $z_s = 2.0$ (Figure 4).

In addition to the frequency of multiple deflections, 
Figures 2, 3, and 4 make an important point about the role of 
the cosmological parameters in galaxy-galaxy lensing.  By and large, the number and
magnitude of individual weak lensing deflections is unaffected by the choice
of the cosmological parameters.  That is, galaxy-galaxy lensing primarily
provides information about the potentials of the lens galaxies, not the cosmology
per se (see also BBS).  Therefore, for the remainder of the manuscript a flat
$\Lambda$-dominated universe with $H_0 = 70$~km~sec$^{-1}$~Mpc$^{-1}$,
$\Omega_{m0} = 0.3$, and $\Omega_{\Lambda 0} = 0.7$ will be adopted.

While galaxy-galaxy lensing is largely insensitive to the values of the
cosmological parameters, it is quite sensitive to masses of the halos
of the lens galaxies.
The dependence of galaxy-galaxy lensing on the physical radii of the halos 
of the lens galaxies
is rather weak (see, e.g., BBS; Hoekstra et al.\ 2004; Kleinheinrich et al.\ 2006); 
however, the dependence of galaxy-galaxy lensing
on the velocity dispersions of the halos of the lens galaxies is quite strong.
The effect of varying the characteristic lens parameters on
the frequency of multiple weak deflections is shown in Figures 5, 6, and 7.
In contrast with Figures 2, 3, and 4,
here the source galaxies have been distributed broadly in redshift space (as in
Figure 1), with a median source redshift of $z_s \sim 0.96$. 
The characteristic
halo parameters for $L_B^\ast$ galaxies are varied as follows: 
$\sigma_v^\ast = 135$~km~sec$^{-1}$, $\sigma_v^\ast = 150$~km~sec$^{-1}$,
$\sigma_v^\ast = 165$~km~sec$^{-1}$; $s^\ast = 50h^{-1}$~kpc, 
$s^\ast = 100h^{-1}$~kpc, $s^\ast = 200h^{-1}$~kpc.  As in Figures 2, 3, and 4,
the shear produced by each individual deflection is restricted to
$\gamma > 0.0025$ (Figure 5), $\gamma > 0.005$ (Figure 6), and $\gamma > 0.01$
(Figure 7).  

\begin{figure}
\centerline{\scalebox{0.70}{\includegraphics{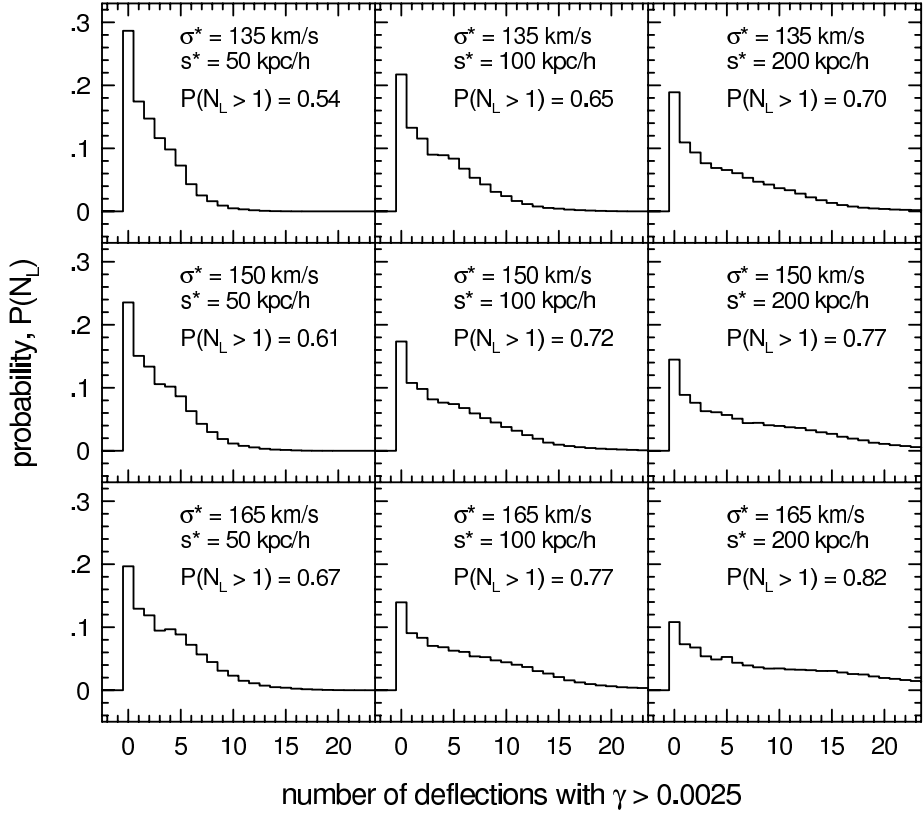} } }%
\vskip -0.0cm%
\caption{
Probability, $P(N_L)$, that a source galaxy has been lensed by
$N_L$ foreground galaxies, where each individual
lens induces a shear $\gamma > 0.0025$.
For $N_L > 1$, multiple deflections with $\gamma > 0.0025$ have
been experienced by the source.  Here the sources have been
distributed broadly in redshift space with a median
redshift $z_s = 0.96$, and a flat $\Lambda$-dominated universe
with $H_0 = 70$~km~sec$^{-1}$~Mpc$^{-1}$, $\Omega_{m0} = 0.3$ and
$\Omega_{\Lambda 0} = 0.7$
has been adopted.  Lens galaxies have a median redshift
$z_l = 0.55$.  Different panels correspond to different characteristic
parameters ($\sigma_v^\ast$, $s^\ast$) adopted for the 
halos of $L_B^\ast$ lens galaxies.
}
\label{fig5}
\end{figure}

\begin{figure}
\centerline{\scalebox{0.70}{\includegraphics{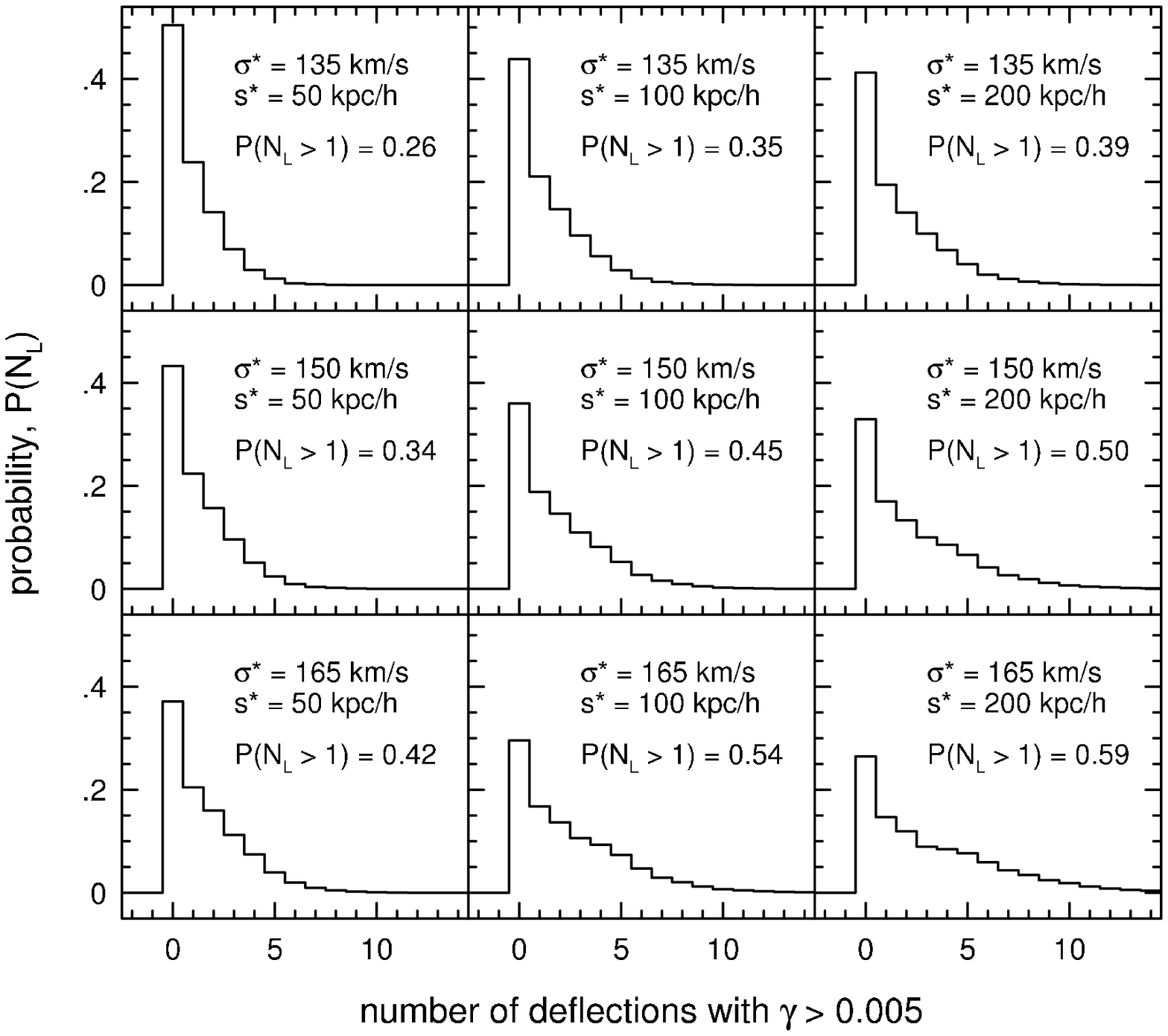} } }%
\vskip -0.0cm%
\caption{
Same as Figure 5, except here the frequency of deflections with
$\gamma > 0.005$ is shown.  For $N_L > 1$, multiple deflections
have been experienced by the source.
}
\label{fig6}
\end{figure}

\begin{figure}
\centerline{\scalebox{0.70}{\includegraphics{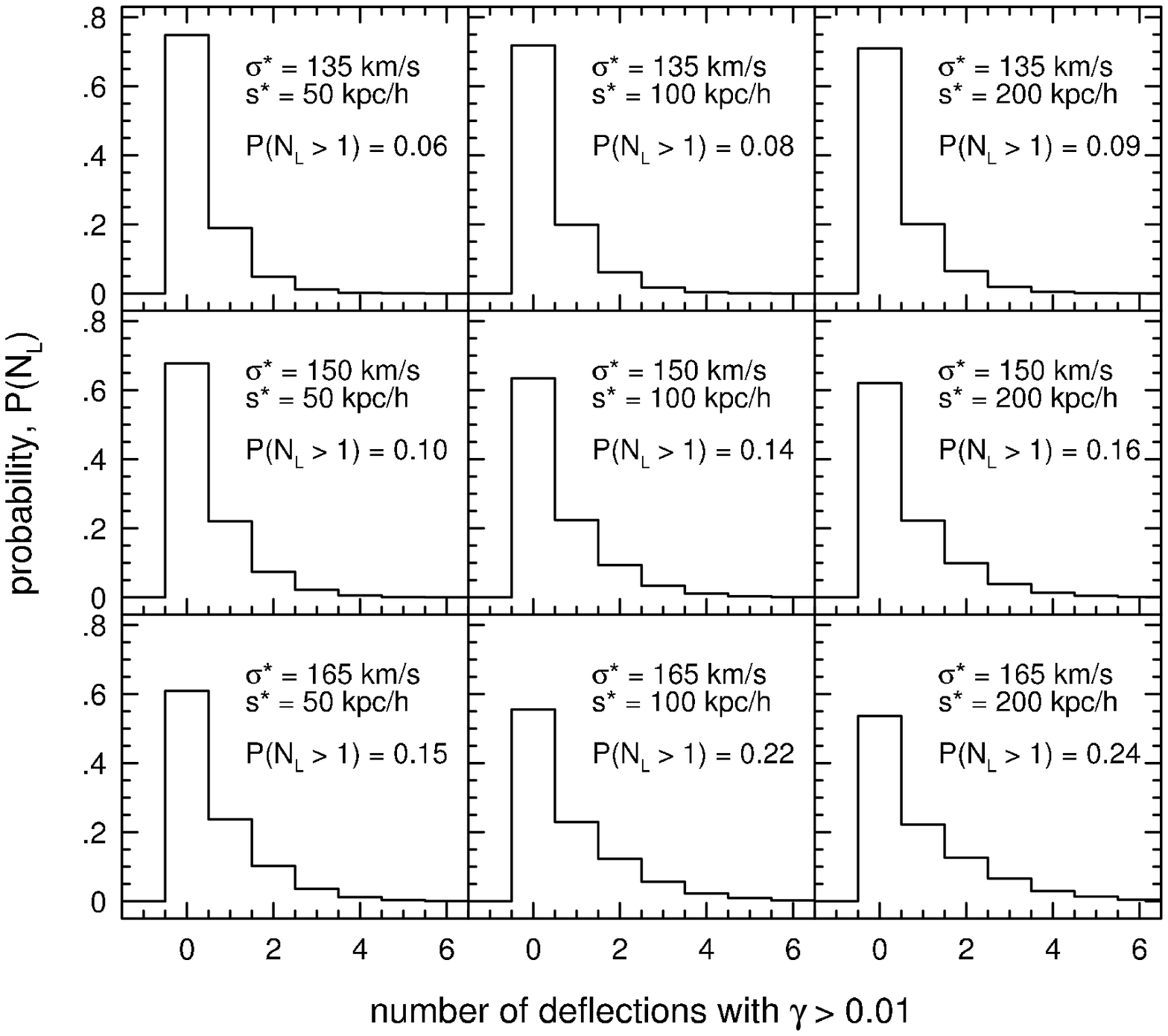} } }%
\vskip -0.0cm%
\caption{
Same as Figure 5, except here the frequency of deflections with
$\gamma > 0.01$ is shown.  For $N_L > 1$, multiple deflections have
been experienced by the source.
}
\label{fig7}
\end{figure}

For the adopted source redshift distribution, then, the probability of
multiple weak deflections increases as the characteristic mass of the halos
of $L_B^\ast$ lens galaxies increases.  That is, the larger is the mass of the
lens, the wider is its aperture of influence on the sky.  For the adopted 
source redshift distribution, there is a high probability of multiple deflections in which
each individual deflection results in a shear of $\gamma > 0.0025$.
The probability ranges from 54\% for the lowest characteristic halo mass (Figure
5, top left) to 82\% for the highest characteristic halo mass (Figure 5, bottom
right).  Similarly, there is a high probability of multiple deflections in which each
individual deflection results in a shear of $\gamma > 0.005$.  The
probability ranges from 26\% for the lowest characteristic halo mass (Figure
6, top left) to 59\% for the highest characteristic halo mass (Figure 6, bottom
right).  From Figure 7, instances of multiple deflections in which each individual
deflection results in a very substantial shear of $\gamma > 0.01$ are relatively
rare for low values of $\sigma_v^\ast$ and $s^\ast$.  However, for large values
of $\sigma_v^\ast$ and $s^\ast$, the probability can exceed 20\%.
Note that, at fixed impact parameter, the deflection angle,
$\alpha$, caused by $L_B^\ast$ lenses
scales as essentially $s^\ast (\sigma_v^\ast)^2$. 
So, for a lens
with a given velocity
dispersion, the deflection angle scales approximately
linearly with $s^\ast$.  This naturally leads to larger induced
shear for larger values of
$s^\ast$, and a correspondingly larger number of individual deflections that
exceed the minimum shear thresholds used in Figures 5-7. 

\section{Multiple Deflections vs.\ Single Deflections}

The previous section explored the frequency with which source galaxies undergo
multiple weak deflections in a deep galaxy-galaxy lensing data set.  This section
will explore how the net shear, $\gamma_{\rm net}$, obtained from a full,
multiple-deflection calculation compares to the shear obtained solely from the
closest lens in projection on the sky ($\gamma_{\rm close}$), as well as how
the net shear compares to the shear resulting from the largest individual
deflection in the multiple-deflection calculation ($\gamma_{\rm max}$).  In 
particular, the following questions will be addressed:

\begin{itemize}
\item 
Is the closest weak
lens (in projection on the sky) necessarily the most important weak lens?
\item
Is the net shear for a given source galaxy in the multiple-deflection
calculation larger or smaller than the shear induced by the closest lens?
\item
Is the net shear for a given source galaxy in the multiple-deflection
calculation larger or smaller than the shear resulting from the largest
individual weak deflection?
\item
What effect does the inclusion of multiple deflections have on 
the mean tangential shear measured about the lens centers?
\end{itemize}

Throughout this section, source galaxies in the Monte Carlo simulations
will be taken to have the broad redshift distribution used in Figure 1 
(e.g., equation 6 above) and a flat $\Lambda$-dominated universe will be
used.  

The first question of this section is addressed in Figure 8. Here the 
probability distribution for the distance between the strongest
individual weak lens, $\theta_{\rm max}$, and the closest weak lens,
$\theta_{\rm min}$, is shown.
The different panels correspond to 
different characteristic halo parameters that have been adopted for 
the halos of $L_B^\ast$ galaxies (and appropriately scaled for all lenses according
to equations 2 and 3 above).
Figure 8 shows that, in general, the closest lens in projection on the sky
is not the strongest individual weak lens.  That is, of order 50\% of the time
the closest lens is
not the ``most important'' weak lens.  
Figure 8 also shows the importance of performing the multiple-deflection calculation
using sources that are contained within an area that is smaller than
the area covered by the lenses, since the angular separation between the
closest lens to a given source and the most important lens for that 
same source can
reach scales of more than 2 arcminutes.  In particular,
$\sim 35$\% of the strongest lenses
have angular separations $\gtrsim 20$~arcsec from the sources and
$\sim 10$\% of the strongest lenses have angular separations 
$\gtrsim 60$~arcsec from the sources.

\begin{figure}
\centerline{\scalebox{0.70}{\includegraphics{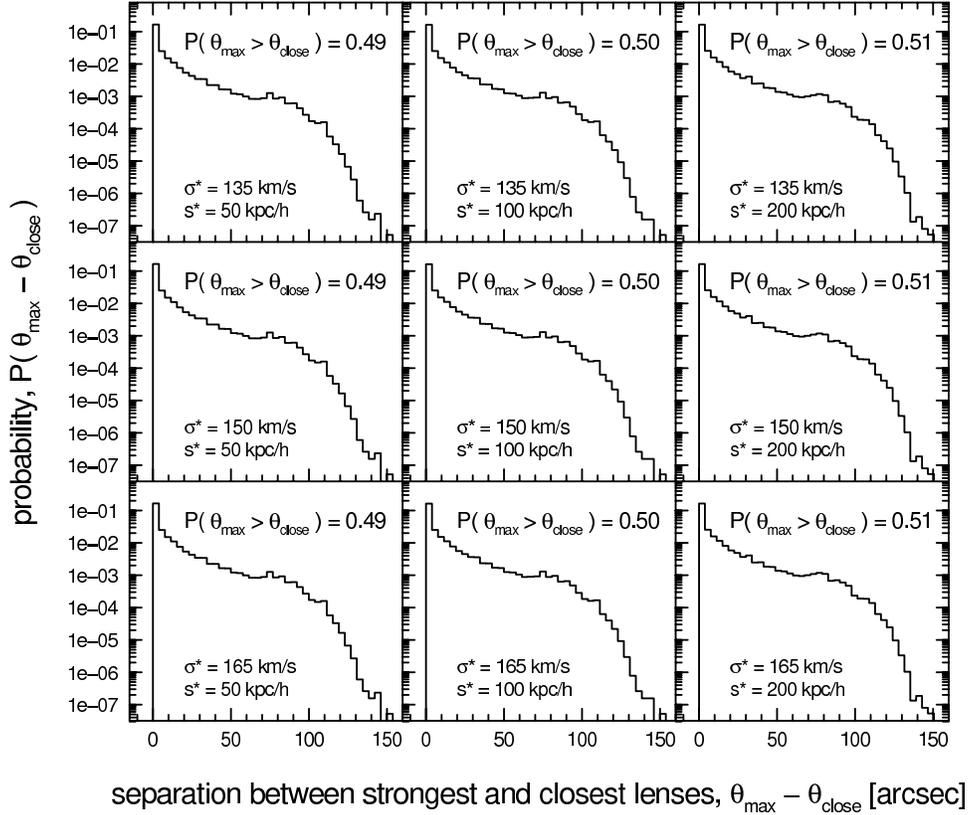} } }%
\vskip -0.0cm%
\caption{
Probability distribution for the distance between the
strongest individual weak lens for a given source, $\theta_{\rm max}$,
and the closest individual weak lens for a given source, $\theta_{\rm close}$.
The distance is zero when the closest lens is, in fact, the strongest lens
for a given source.  The probability that the strongest individual
lens for a given
source is not the closest lens is given in each panel, and is of order
50\% in all cases.
Different
panels correspond to different characteristic halo parameters 
($\sigma_v^\ast$, $s^\ast$) adopted for $L_B^\ast$
lens galaxies. 
}
\label{fig8}
\end{figure}

Figures 9 and 10 address the second and third questions of this section.  That is,
how does the net shear experienced by source galaxies in a full, multiple-deflection
calculation compare to the shear due
to only the closest lens (Figure 9) and to the shear due to the strongest individual
weak lens (Figure 10)?  Figure 9 shows that the net shear 
due to all foreground lenses is generally larger than the shear 
induced by the closest lens on the sky.  The ratio of the shears, $\gamma_{\rm net} /
\gamma_{\rm close}$, is weakly-dependent upon the specifics of the lens halo parameters.
The probability that the net shear in the full multiple-deflection calculation exceeds
the shear due to the single closest lens ranges from 78\% (lens halos with small
physical extents, $s^\ast = 50h^{-1}$~kpc) to 82\% (lens halos with large physical
extents, $s^\ast = 200h^{-1}$~kpc).  
Figure 10 shows that the net shear due to all foreground lenses
is also generally larger than the shear induced
by the strongest individual weak lens in the full, multiple-deflection calculation.
As in Figure 9, the ratio of the shears, $\gamma_{\rm net} / \gamma_{\rm max}$,
is weakly-dependent upon the specifics of the lens halo parameters.  The probability
that the net shear in the full multiple-deflection calculation exceeds the
shear due to the single strongest weak lens ranges from 69\% (lens halos with
small physical extents, $s^\ast = 50h^{-1}$~kpc) to 76\% (lens halos with
large physical extents, $s^\ast = 200^{-1}$~kpc).  Figures 9 and 10, then, show that
for any given distant source galaxy, the net shear that its image experiences due
to all foreground lenses exceeds the shear due solely to the closest lens,
as well as the shear due solely to the strongest individual weak lens.  

\begin{figure}
\centerline{\scalebox{0.70}{\includegraphics{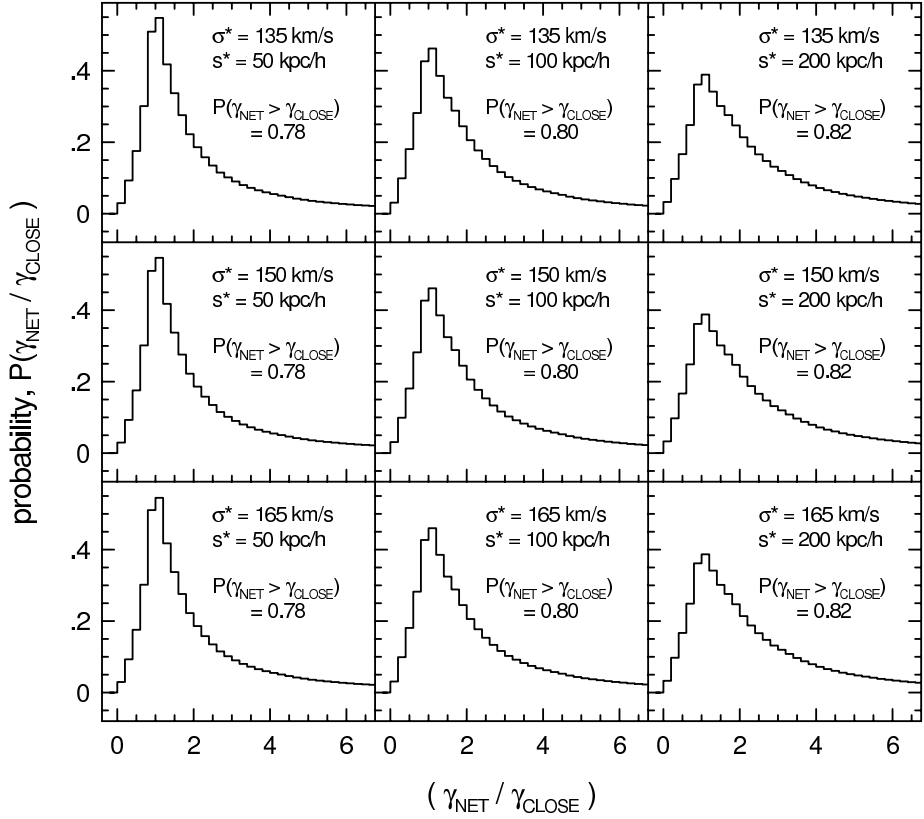} } }%
\vskip -0.0cm%
\caption{
Probability distribution for the ratio of the net shear experienced by
the images of source galaxies, $\gamma_{\rm net}$, to the shear induced solely 
by the closest lens on the sky, $\gamma_{\rm close}$.  Different panels correspond
to different characteristic halo parameters ($\sigma_v^\ast$, $s^\ast$) 
adopted for $L_B^\ast$ lens galaxies.  The
probability that $\gamma_{\rm net}$ exceeds $\gamma_{\rm close}$ is listed in each
panel.
}
\label{fig9}
\end{figure}

\begin{figure}
\centerline{\scalebox{0.70}{\includegraphics{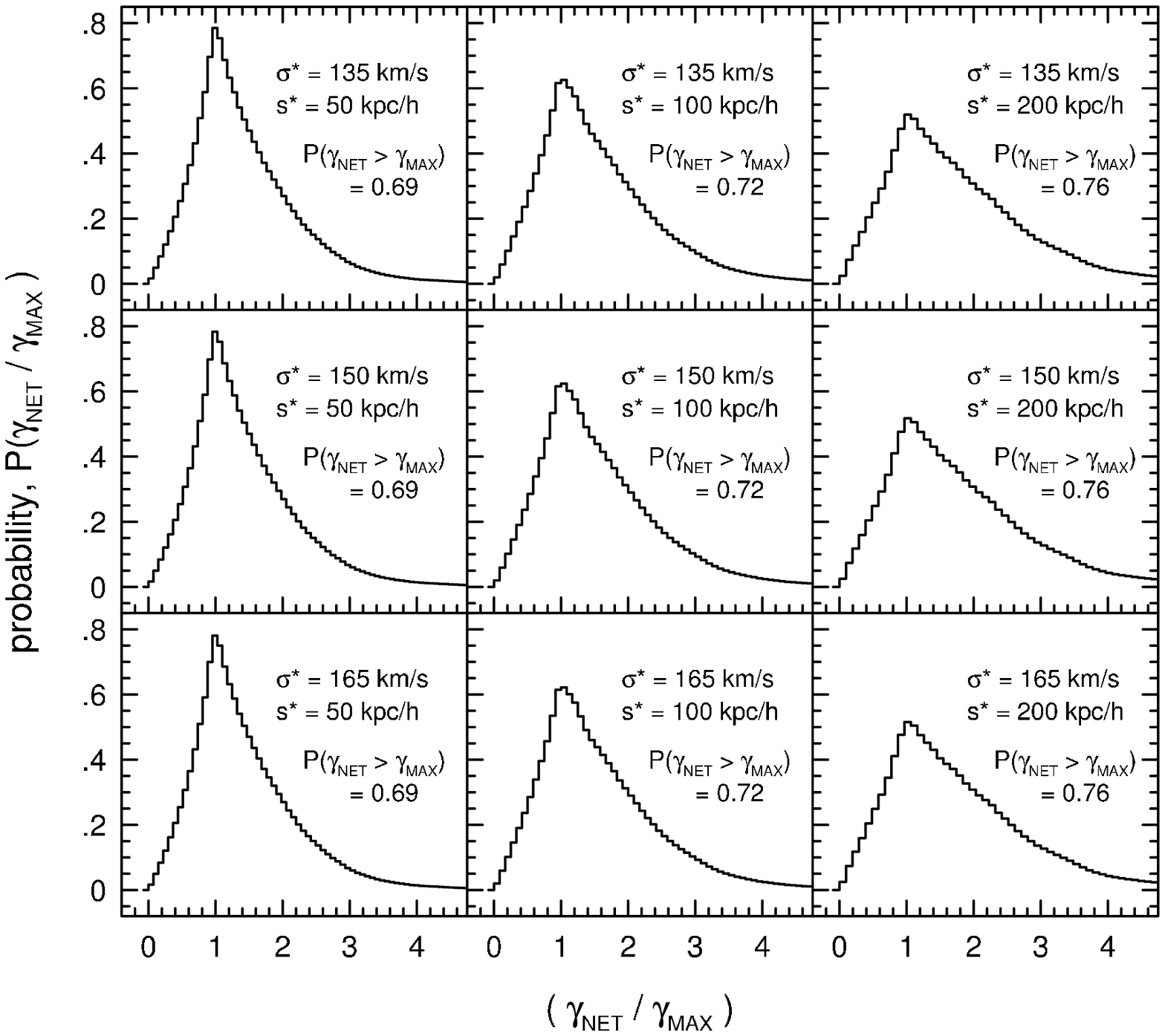} } }%
\vskip -0.0cm%
\caption{
Same as Figure 9, except here the net shear, $\gamma_{\rm net}$, is compared to the
shear induced by the strongest individual lenses in the multiple-deflection 
calculation, $\gamma_{\rm max}$.
}
\label{fig10}
\end{figure}

It may seem somewhat counter-intuitive that
the net shear experienced by the images of distant source galaxies
in the multiple-deflection calculations
generally exceeds the shear due to a naive single-deflection calculation.
That is, at first glance
one might expect that multiple weak galaxy-galaxy lensing deflections should, on average,
cancel each other.  For a given source this would, indeed, be the case if all the 
foreground lenses were located
at precisely the same angular separation from the source, had identical gravitational
potentials, and had identical redshifts, $z_l$.  Such an idealized situation
is, of course, not the case in
the real universe.  
That is, we cannot think in terms of a single lens plane 
for the galaxy-galaxy lensing problem, and to a certain extent the solution 
has to be understood numerically.
This is due to the fact that there are a wide range of lens-source separations, 
the lenses have a wide range of gravitational potentials,
and the lenses 
are distributed broadly in redshift.  
These, in combination, result in increased shear in the multiple-deflection 
calculation for galaxy-galaxy lensing,
much as the non-uniformities in the mass distribution along the 
line of sight give rise to a net ``cosmic shear'' (see, e.g., reviews by
Bartelmann \& Schneider 2001; van Waerbeke \& Mellier 2003;
Refregier 2003; Munshi et al.\ 2008).  That is, like
galaxy-galaxy lensing, cosmic shear is inherently a multiple-deflection problem
in which the deflections do not simply cancel.   A detailed investigation of
how the shear experienced by a given source galaxy is affected as one successively
adds in more and more weak galaxy lenses will be presented in Howell \& Brainerd (2010).

The last question of this section, the effect of multiple deflections on the mean
tangential shear about the lens centers, is addressed in Figure 11.  In Figures 9 and 10,
we have computed quantities (net shear, shear due to the closest weak
lens, and shear due to the
strongest individual weak
lens) that cannot, in practice, be measured in an observational
data set.  That is, without precise knowledge of the intrinsic shape of a source galaxy,
the angular diameter distances of the source and all possible foreground lens galaxies, 
as well as
the details of the gravitational potentials of all foreground lens galaxies, it is not
possible to deduce $\gamma_{\rm net}$, $\gamma_{\rm close}$, and $\gamma_{\rm max}$ for any
one source galaxy.  Indeed, galaxy-galaxy lensing yields such a small value of 
$\gamma_{\rm net}$ that it can only be detected via an ensemble average over the images of
many source galaxies.  Therefore, Figure 11 demonstrates the effect of 
multiple deflections on the {\it observable} galaxy-galaxy lensing signal: the mean tangential
shear about the lens centers.

Shown in Figure 11 is the mean tangential shear, $\gamma_T (\theta)$,
measured as a function of lens-source
angular separation.  Since the sources are restricted to a circle of
radius 2.5~arcmin, while the lenses are restricted to a circle of
radius 4.0~arcmin, it is not possible to compute $\gamma_T (\theta)$ around
all of the lenses.  Instead, $\gamma_T (\theta)$ is computed using only those lenses
that are within a distance $r = (150-\theta)$~arcsec of the center of the field.
This allows the average to be computed in complete circular annuli, centered on each
lens galaxy, and avoids edge effects.  
Solid squares in Figure 11 show
$\gamma_T (\theta)$ for the full multiple-deflection calculations in which each
source has been lensed by all foreground lenses.
Open circles in Figure 11 show $\gamma_T (\theta)$ for single deflection
calculations in which each source is lensed by only the closest lens on the sky.
Shown in Figure 12 is the ratio of the mean tangential shears that are
plotted in Figure 11.  That is, Figure 12 shows the ratio of the mean tangential
shear obtained from the full multiple-deflection calculations to that obtained
from the single-deflection calculations.

\begin{figure}
\centerline{\scalebox{0.70}{\includegraphics{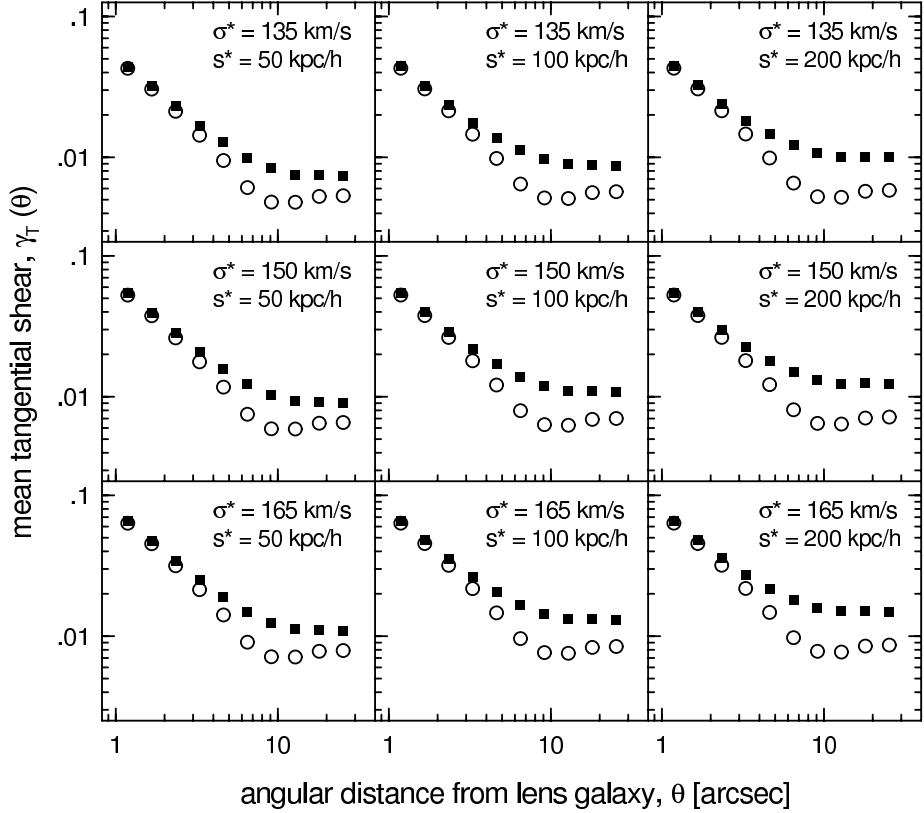} } }%
\vskip -0.0cm%
\caption{
Mean tangential shear, $\gamma_T (\theta)$, computed in circular annuli of
radius $\theta$, centered on the lens galaxies. 
Different panels correspond to different characteristic parameters
($\sigma_v^\ast$, $s^\ast$) adopted
for the halos of $L_B^\ast$ lens galaxies.  Solid squares: results of full
multiple-deflection calculations in which source galaxies have been lensed by all
foreground galaxies.  Open circles: results of
single-deflection calculations in which source galaxies are lensed only by the
closest lens.  The mean angular separation between the lenses is $\theta = 10.7''$.
}
\label{fig11}
\end{figure}

\begin{figure}
\centerline{\scalebox{0.70}{\includegraphics{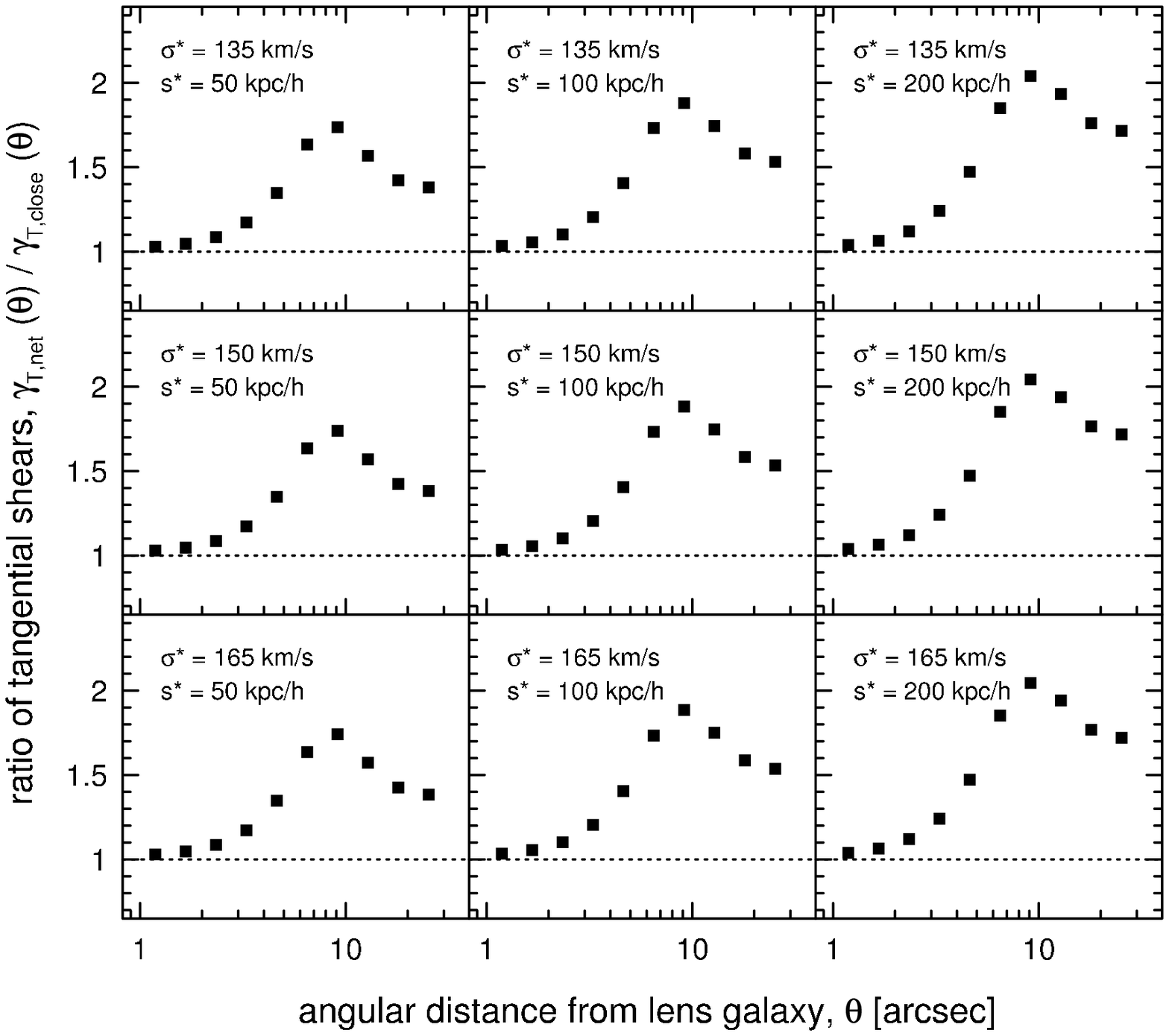} } }%
\vskip -0.0cm%
\caption{
Ratio of the mean tangential shears shown in Figure 11.  Different panels correspond
to different characteristic parameters ($\sigma_v^\ast$, $s^\ast$), adopted for
the halos of $L_B^\ast$ lens galaxies.  Dotted line indicates a value of unity. 
On scales of $\theta \gtrsim 2$~arcsec
the mean tangential shear from the full multiple deflection 
calculations, $\gamma_{\rm T,net} (\theta)$, exceeds the mean tangential 
shear from the single-deflection calculations, $\gamma_{\rm T, close} (\theta)$.
}
\label{fig12}
\end{figure}

From Figures 11 and 12, then, it is clear that on very small scales, galaxy-galaxy
lensing reduces to a single-deflection problem.  That is, on scales $\theta \sim
1$~arcsec, there is relatively little difference between the mean tangential shear obtained
from the full, multiple-deflection calculations and the single deflection calculations.
On scales of $\theta \gtrsim 2$~arcsec, however, the
multiple deflection calculations yield a higher value of the mean tangential shear.
The difference between $\gamma_T(\theta)$ from the multiple deflection calculations
and $\gamma_T(\theta)$ from the single deflection calculations depends somewhat on
the characteristic parameters adopted for the halos of $L_B^\ast$ galaxies.  
On scales
$\theta \sim 20$~arcsec, the multiple deflection calculation for the 
lowest mass halos (top left panel of Figure 12) yields
a mean tangential shear that is a factor of $\sim 1.4$ larger than the mean tangential shear
from the single deflection calculation.   The multiple deflection calculation for
the highest mass halos (bottom right panel of Figure 12)
yields a mean tangential shear that is a factor of $\sim 1.7$ larger than the mean tangential
shear from the single deflection calculation for $\theta \sim 20$~arcsec.
It is also interesting to note that $\gamma_T (\theta)$ becomes roughly constant on scales
$\theta \gtrsim 10$~arcsec.
This is due to the fact that the mean angular separation
between the lens galaxies is 10.7~arcseconds, which corresponds to a comoving distance
of $48 h^{-1}$~kpc at the median redshift of the lenses.  That is, 
on angular scales comparable to and larger than the mean angular
separation of the lenses, the halos of nearby lens galaxies are overlapping
one another in projection on the sky.  This is the primary reason that galaxy-galaxy
lensing is not terribly sensitive to the radii of the halos of the lens galaxies.

It should be kept in mind that here we have only modeled the masses of 
the halos of individual galaxies.  In particular, we have not included
the fact that many of the larger, brighter galaxies are probably contained within
group environments that have a substantial dark matter component over and
above the dark matter halos of the individual galaxies.  
The circularly-averaged
tangential shear (as we have computed here) is related to the surface mass density
of a circular lens through
\begin{equation}
\gamma_T \Sigma_c = \overline{\Sigma}(< \theta) - \overline{\Sigma}(\theta) \equiv
\Delta \Sigma
\end{equation}
(e.g., Miralda-Escud\'e 1991).  In the case of completely 
isolated lens galaxies, $\Delta \Sigma$
above is the surface mass density of the dark matter halo of the lens galaxy.  In
the case of lenses that reside within groups and clusters, $\Delta \Sigma$ includes
the mass due to the dark matter halos of the lens galaxies, as well as the mass
of the larger
dark matter halo that surrounds the group or cluster.
In the case of bright, massive galaxies that reside
in groups and clusters, 
the tangential shear shown in Figure 11 does not properly correlate with all
of the mass that one would actually
expect to contribute to the net shear in a observational
data set.  This is simply because we have neglected the additional mass associated with
the environments in which those galaxies tend to reside. 

Observations of galaxy-galaxy lensing have shown that the dependence of the
tangential shear on projected distance from the lens is a function of the
stellar mass and luminosity of the lens.  In particular, 
the tangential shear measured around lens galaxies with low
stellar masses and low luminosities is approximately constant
at very large projected distances (e.g., Mandelbaum et al. 2006a, Figures 1 and 2), 
consistent with
the results in Figure 11 above.  However, the
tangential shear measured around lens galaxies with high stellar masses and high
luminosities declines monotonically at large projected distances
(e.g., Mandelbaum et al. 2006a, Figures 1 and 2).
This is due to the contribution of the overall mass within the relatively higher
density environments in which the most massive, most luminous galaxies
tend to reside.  That is, in practice observed galaxy-galaxy lensing includes 
the effects of all individual galaxy lenses, as well as the effects of
the mass within the local environment that surrounds the lenses.
Here we have simply considered the effects of the individual halos of bright 
galaxies, and have not included environmental (e.g., group/cluster)
contributions to the net shear.

The implications of Figures 11 and 12 are straightforward.  If one wishes to use observations
of $\gamma_T (\theta)$ to constrain the fundamental parameters associated with the
halos of the lens galaxies (i.e., $\sigma_v^\ast$ and $s^\ast$ for the model 
adopted here), it is vital to use full, multiple-deflection Monte Carlo simulations
for the parameter fitting.
If simple, single-deflection calculations are used, the inferred halo masses will
be systematically too large.  That is, in order to reproduce an observed galaxy-galaxy
lensing signal on angular scales greater than a few arcseconds using a single-deflection
calculation, one would need systematically larger halo masses than are required in
the full multiple-deflection calculation.

\section{Galaxy-Galaxy Lensing and Cosmic Shear}

The galaxy-galaxy lensing contribution
to cosmic shear is investigated in this section.  
Cosmic shear is often equated to weak lensing
by the large-scale structure of the universe, but in practice cosmic shear is the
result of photons from distant source galaxies being deflected by all mass along
the line of sight.  The mass along the line of sight
 includes large galaxy clusters, galaxy groups, and filaments,
as well as objects with smaller masses such as individual galaxies.  
In the case of the galaxy-galaxy lensing,
we are considering the specific contribution of the
highly non-linear, large $k$ 
contribution of the power spectrum of density fluctuations, $P(k)$, to the cosmic
shear signal.

The source galaxies in the Monte Carlo simulations are assumed to have orientations
that are intrinsically uncorrelated (i.e., each source is assigned an
initially random position angle).  Galaxy-galaxy lensing will, of course, slightly change
both the ellipticity and the orientation of each source.  In the presence of a
number of high-mass lens galaxies that cause multiple weak deflections over large
angular scales, the images of the source galaxies may acquire a net preferred
orientation due to galaxy-galaxy lensing.  This is the signature of cosmic shear, 
albeit in this case the shear is caused solely by the lens galaxies, not the 
entire large-scale
structure of the universe.

To investigate the degree to which galaxy-galaxy lensing may contribute to the
cosmic shear signal, the image correlation function
\begin{equation}
C_{\chi \chi} (\theta) = \left< \vec{\chi}_{f,i} \cdot \vec{\chi}_{f, j}^\ast \right>_\theta
\end{equation}
is computed.  Here the mean is computed over all galaxy pairs $i, j$ separated by an 
angle $\theta \pm d\theta/2$, $\vec{\chi}_{f,i}$ is the final shape parameter of
source galaxy $i$, and $\vec{\chi}_{f,j}^\ast$ is the complex conjugate of the 
final shape parameter of galaxy $j$ (see, e.g., Blandford et al.\ 1991).  The 
correlation function measures the extent to which galaxy images ``point'' in the same
direction on the sky.  If $C_{\chi \chi}(\theta)$ is positive, the images of the galaxies
are aligned with each other.  If $C_{\chi \chi}(\theta)$ is zero, the images of the
galaxies are randomly oriented.  If $C_{\chi \chi}(\theta)$ is negative, the images of
the galaxies tend to be oriented perpendicular to each other (i.e., they are anti-aligned).

Shown in Figure 13 is the image correlation function for the source galaxies,
where the galaxies have again been broadly distributed in redshift space and
a flat $\Lambda$-dominated universe has been adopted.  Solid squares show the results
for the full, multiple-deflection calculations and open circles show the results
for the single-deflection calculations in which the sources have been lensed by 
only the closest lenses.  From Figure 13, then, it is clear that for angular separations
$\theta \gtrsim 5$~arcsec, the single-deflection calculations yield essentially
no contribution of galaxy-galaxy lensing to the cosmic shear.  That is, if multiple
deflections were not important in galaxy-galaxy lensing, one would expect that on
scales greater than 5~arcsec, galaxies alone would not contribute to cosmic shear.
Hence, cosmic shear on scales greater than 5~arcsec would be expected to be
largely independent
of the gravitational potentials of the halos of field galaxies.
However, the full, multiple-deflection calculations in Figure 13
show that galaxy-galaxy lensing can, indeed, induce substantial correlations
in the source images on
scales greater than 5~arcsec.  Furthermore, the degree of lensing-induced
image alignment is strongly affected by
the characteristic parameters that are adopted for the halos of $L_B^\ast$ lenses.

\begin{figure}
\centerline{\scalebox{0.70}{\includegraphics{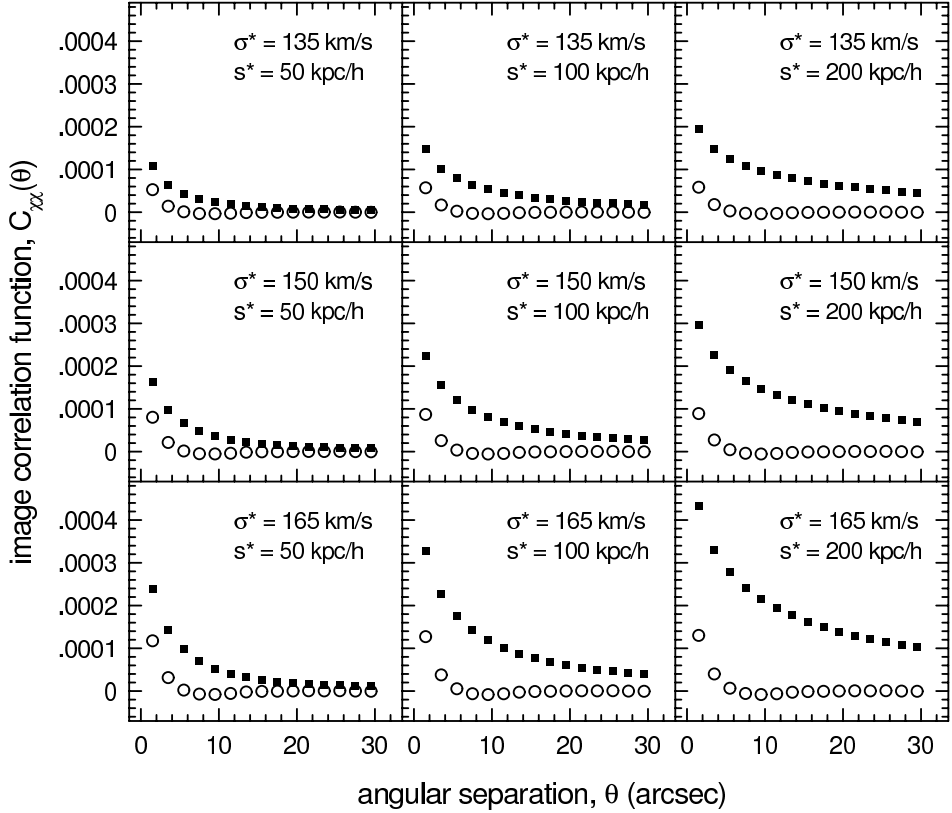} } }%
\vskip -0.0cm%
\caption{
Image correlation function, $C_{\chi\chi}(\theta)$, due to galaxy-galaxy lensing
alone.  Different panels correspond to different characteristic parameters 
($\sigma_v^\ast$, $s^\ast$) adopted
for the halos of $L_B^\ast$ galaxies.
Solid squares: results of full
multiple-deflection calculations in which source galaxies have been lensed by all
foreground galaxies.  Open circles: results of
single-deflection calculations in which source galaxies are lensed only by the
closest lens. 
}
\label{fig13}
\end{figure}

In addition to the image correlation function, the top hat shear variance
\begin{equation}
\left< \gamma^2 \right> = \frac{2}{\pi \theta^2}\int_0^\infty \frac{dk}{k}~
P_\kappa(k) \left[ J_1 (k\theta) \right]^2
\end{equation}
is common measure of cosmic shear. 
Here  $P_\kappa$ is the power spectrum of the projected mass density
of the universe, $J_1$ is a Bessel function of the first kind, and
$\theta$ is the radius of the circular aperture over which the mean is computed.
In an observational data set, the function is computed as
\begin{equation}
\left< \gamma^2 \right> = 
\frac{1}{N (N-1)} \sum_{i\ne j} \vec{\gamma}_i \cdot \vec{\gamma_j}^\ast,
\end{equation}
for all galaxies within a circular aperture of radius $\theta$ on the sky.
Solid squares in Figure 14 show the shear top hat variance due to galaxy-galaxy lensing
alone, obtained from full, multiple-deflection calculations.  Again, sources
have been broadly distributed in redshift and a flat $\Lambda$-dominated universe is
adopted.  Also shown for comparison (crosses connected by dotted line) are the
measured values of $\left < \gamma^2 \right>$ obtained by Fu et al.\ (2008) for
galaxies in the CFHT Legacy Survey with a median redshift $z_m = 0.83$.   Although
this is somewhat lower than the median redshift of the source galaxies in the
Monte Carlo simulations, it is sufficiently similar that it is reasonable to compare
the observational and theoretical results directly.

From Figure 14, the small-scale contribution of galaxy-galaxy lensing to
$\left< \gamma^2 \right>$
depends quite strongly on the parameters adopted for the halos of $L_B^\ast$ galaxies,
and scales roughly with the relative masses of the halos.
For example, the lowest mass $L_B^\ast$ lenses (top left panel) have masses that are a
factor of 2.5 smaller than those of the fiducial halo with $\sigma_v^\ast =
150$~km~sec$^{-1}$ and $s^\ast = 100h^{-1}$~kpc (middle panel).
Similarly, the highest mass $L_B^\ast$ lenses (bottom right panel) have masses that
are a factor of 2.4 larger than those of the fiducial halo.  At $\theta = 1$~arcmin,
$\left< \gamma^2 \right>$ for the lowest mass lenses is a factor of 3 smaller than
it is for the fiducial halo, and $\left< \gamma^2 \right>$ for the highest mass
lenses is a factor of 3.5 larger than it is for the fiducial halo.

Comparing the squares in Figure 14 (simulation results) to the crosses (observational
results), it is clear that, depending upon how deep the potential wells of 
$L_B^\ast$ galaxies are, galaxy-galaxy lensing alone may contribute a substantial
amount to cosmic shear.  In the case of the lowest mass $L_B^\ast$ halos, galaxy-galaxy
lensing alone would be expected to contribute only $\sim 5.5$\% of the value of 
$\left< \gamma^2 \right>$ measured by Fu et al.\ (2008) for an aperture of radius
$\theta = 1$~arcmin.  In the case of the fiducial $L_B^\ast$ halos, the contribution
of galaxy-galaxy lensing alone increases to $\sim 16$\% for 
$\theta = 1$~arcmin, while for the highest mass
$L_B^\ast$ halos $\sim 58$\% of the signal seen by Fu et al.\ (2008) at $\theta = 1$~arcmin
would be due
to galaxy-galaxy lensing alone.  If one were to extrapolate the Fu et al.\
(2008) results to scales $\theta < 1$~arcmin, the results shown in the
bottom right panel of Figure 14 suggest that the halos of $L_B^\ast$ galaxies are
probably not as large adopted in this particular panel.  That is, a simple extrapolation of
the Fu et al.\ (2008) result to $\theta < 1$~arcmin
leads to an expectation of much less observed cosmic shear
than is predicted by galaxy-galaxy lensing by our highest mass lenses.

\begin{figure}
\centerline{\scalebox{0.70}{\includegraphics{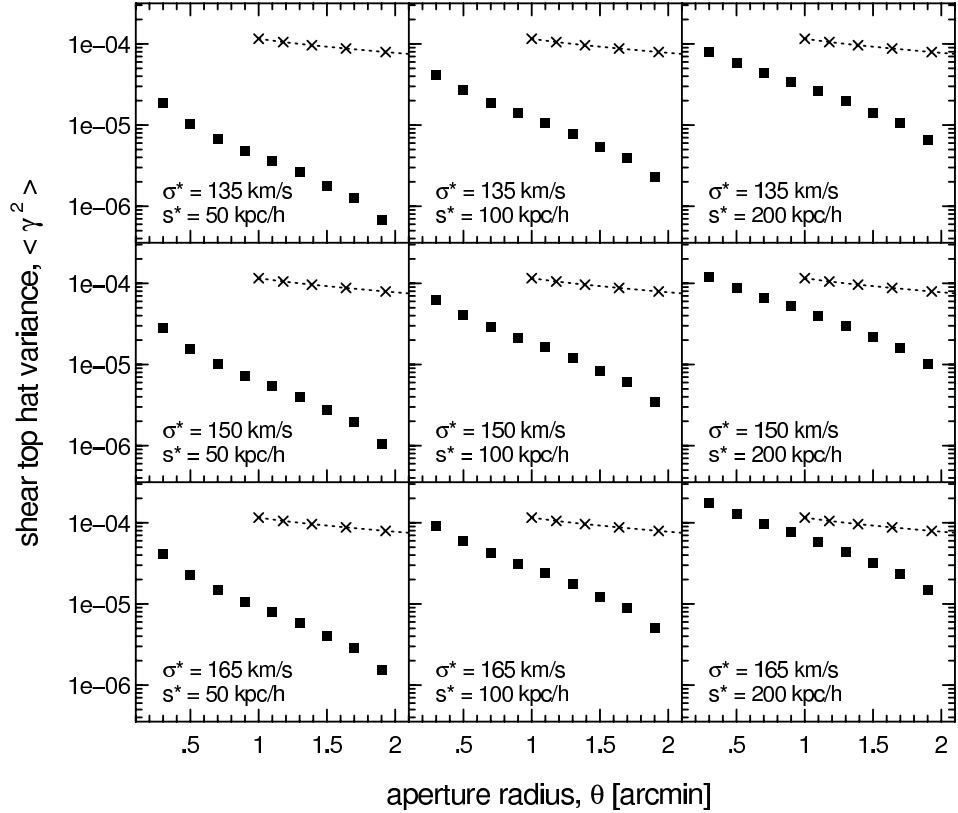} } }%
\vskip -0.0cm%
\caption{
Solid squares: top hat shear variance, $\left< \gamma^2 \right>$, due to galaxy-galaxy lensing
alone, obtained from the full multiple-deflection
calculations.  Different panels correspond to different characteristic
parameters ($\sigma_v^\ast$, $s^\ast$) adopted for the halos of $L_B^\ast$ galaxies.  
Shown for comparison
(crosses connected by dotted line) are the results from Fu et al.\ (2008) for 
the top hat shear variance obtained from the CFHT Legacy Survey using sources 
with median redshift $z_s = 0.83$.
}
\label{fig14}
\end{figure}

Because of the relatively small area of the sky that is covered by the
sources in the Monte Carlo simulations, it is not possible
to compute $\left< \gamma^2 \right>$ on large angular scales.  However, the
rms value, $\left< \gamma^2 \right>^{1/2}$, decreases linearly with $\theta$ and
it is, therefore, straightforward to extrapolate the results from Figure 14 to
an angular scale at which the contribution of galaxy-galaxy lensing to cosmic
shear vanishes.  From this extrapolation, then, the contribution of galaxy-galaxy
lensing to cosmic shear vanishes at $\theta = 5.0$~arcmin for the lowest mass 
halos, $\theta = 5.2$~arcmin for the fiducial model, and $\theta = 5.4$~arcmin for
the highest mass halos.  Therefore, although the contribution of galaxy-galaxy 
lensing to cosmic shear on small angular scales is very sensitive to the details
of the gravitational potentials of galaxies, cosmic shear on scales 
$\theta \gtrsim 5$~arcmin should not be affected by galaxy-galaxy lensing 
to any significant degree.

\section{Cosmic Variance}

The shear field in the Monte Carlo simulations comes from a set of lenses
that are contained within an area of 50 sq.\ arcmin.\ on the sky and, therefore,
one might be concerned that the results shown above could be compromised by
cosmic variance.  Here some of the results above are 
recomputed using subdivisions of the data in order to explore potential
small field effects.  To do this, we use the fiducial model in which the halos
of $L_B^\ast$ galaxies have velocity dispersions of $\sigma_v^\ast = 150$~km~sec$^{-1}$
and characteristic radii of $s^\ast = 100h^{-1}$~kpc, we take the sources
to be broadly distributed in redshift as above, and we use a
flat $\Lambda$-dominated cosmology.

Shown in Figure 15 is a comparison of the frequency of multiple deflections 
that give rise to individual shear values of $\gamma > 0.005$ (top panels),
the mean tangential shear about the lens centers, $\gamma_T (\theta)$ (middle panels), and
the image correlation function, $C_{\chi\chi}(\theta)$ (bottom panels) using
different subdivisions of the data.  The left hand panels show a comparison
of results obtained from data in the northern half of the field and
results obtained from data in the southern half of the filed.  The right hand
panels show results obtained from data in the eastern half of the field and
results obtained from data in the western half of the field.
While there are some differences that result when the size of the
field is reduced by a factor of two, the differences are small and suggest
that the results obtained from the full field should not suffer dramatically
from effects of cosmic variance.

\begin{figure}
\centerline{\scalebox{0.70}{\includegraphics{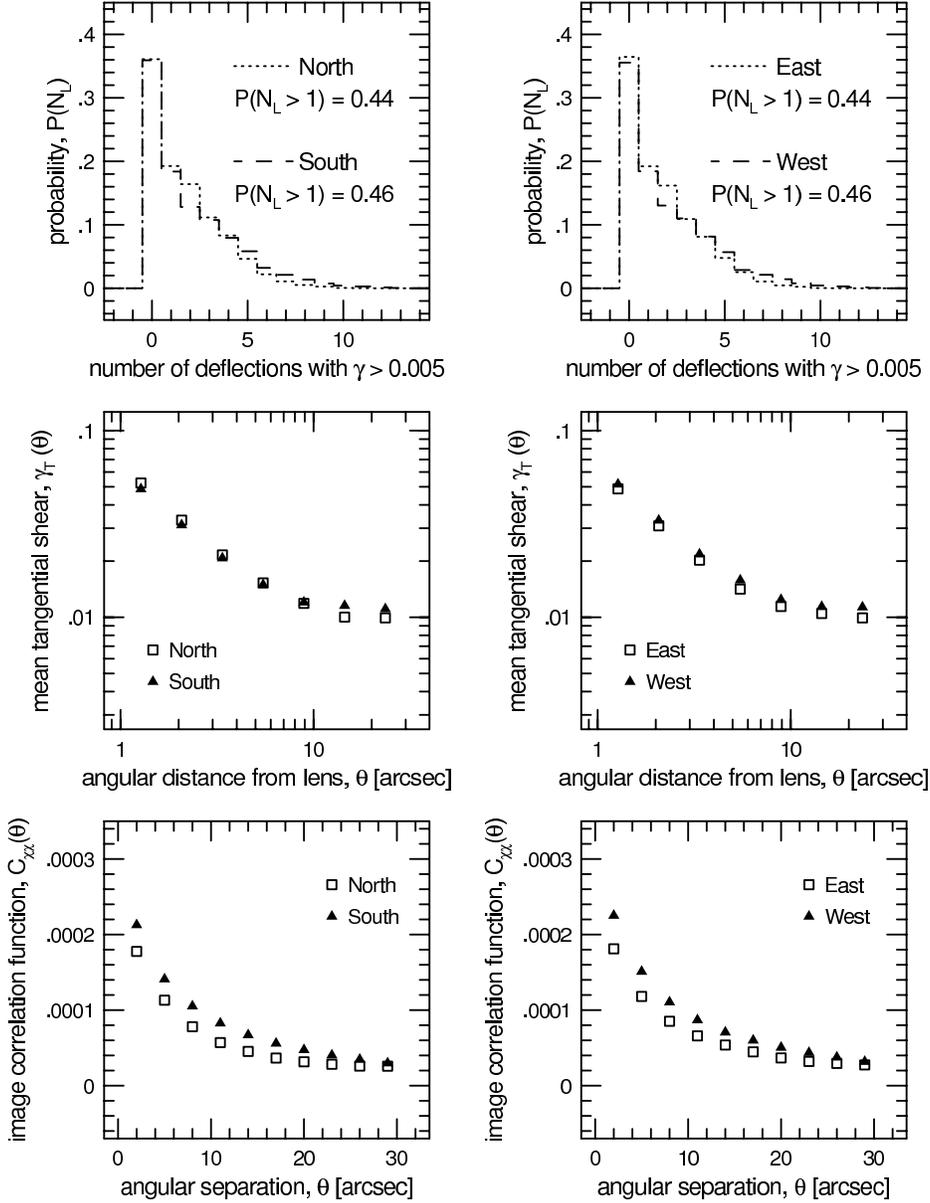} } }%
\vskip -0.0cm%
\caption{
Comparison of the frequency of multiple weak deflections (top),
the mean tangential shear (middle), and the image correlation function
(bottom) for different subdivisions of the field.  Left: comparison
of results from the northern half of the field to results from the
southern half of the field.  Right: comparison of results from the eastern
half of the field to results from the western half of the field.  Here
the sources have been broadly distributed in redshift with $z_{\rm med} = 0.96$.
The fiducial
halo model with $\sigma_v^\ast = 150$~km~sec$^{-1}$ and 
$s^\ast = 100h^{-1}$~kpc, and a flat $\Lambda$-dominated cosmology have also
been adopted.
}
\label{fig15}
\end{figure}

\section{Conclusions}

The frequency and effects of multiple weak lensing deflections in galaxy-galaxy
lensing have been investigated using Monte Carlo simulations.  The lenses in
the simulations are modeled using observed galaxies with magnitudes $R \le 23$,
contained within a circle of radius of 4~arcminutes, centered on the HDF-N.  The lenses have
known redshifts and known rest-frame B-band luminosities.  By adopting a simple
halo mass model it is possible to determine the relative strengths of each of the
lenses using scaling relations.

The Monte Carlo simulations reveal a number of expected results: (i) the 
frequency of multiple deflections depends upon the minimum value of the shear (i.e.,
the lower is the minimum value, the more likely it is that multiple deflections
will be experienced by a given source), (ii) the frequency of multiple
deflections depends upon the source redshift (i.e., the higher is the source
redshift, the more likely it is that will experience multiple deflections) and
(iii) the higher are the masses of the lenses, the more likely it is that
multiple deflections will occur.  For a deep galaxy-galaxy lensing data set
in which the sources have a median redshift $z_s \sim 1$ and the lenses have
a median redshift $z_l \sim 0.6$, the probability that a given source galaxy will
have experienced more than one weak lens that induces a ``typical'' shear
of $\gamma = 0.005$ ranges from 26\% to 59\%, depending upon the masses adopted
for the lenses.  

The Monte Carlo simulations also reveal a number of results that may seem
counter-intuitive at first glance: (i) of order 50\%
of the time, the closest lens in projection on the
sky is not the most important weak lens for a given source, (ii) for a
given source, the net shear
due to all foreground lenses generally exceeds the shear due to the strongest
individual weak lens, and (iii) multiple deflections give rise to a larger tangential
shear around the lens galaxies than a simple, single-deflection calculation
in which the closest lens is assumed to be the only lens.  This emphasizes
the importance of using full, multiple-deflection calculations when using 
observations of galaxy-galaxy lensing to constrain the parameters of the dark
matter halos of the lens galaxies.  If multiple deflections are not incorporated
into the calculation, this will result in halo masses that are systematically 
too large.

Lastly, the Monte Carlo simulations reveal that galaxy-galaxy lensing alone
can give rise to a cosmic shear signal on small angular scales.  This is 
unsurprising because cosmic shear occurs when photons from distant galaxies
are deflected by all mass along the line of sight.  In the case of galaxy-galaxy
lensing, it is the very large $k$ end of the power spectrum of density fluctuations
that contributes to the cosmic shear by inducing correlated image shapes for
the distant galaxies.  On scales $\theta \sim 1$~arcmin, the degree to which
galaxy-galaxy lensing contributes to cosmic shear is quite sensitive to the
masses of the lens galaxies.  Changing the mass of the halo of a fiducial 
$L_B^\ast$ galaxy by a factor of $\sim 2.5$ changes the contribution to the
top hat shear variance, $\left< \gamma^2 \right>$, by a factor of $\sim 3$.
Comparing the theoretical values of $\left< \gamma^2 \right>$ at $\theta = 1$~arcmin
to the value observed by Fu et al.\ (2008) for sources with a similar redshift
distribution, galaxy-galaxy lensing alone could account for as little as
$\sim 5$\% or as much as $\sim 58$\% of the observed value, depending upon the
halo mass for $L_B^\ast$ galaxies.  

While the small-scale contribution of galaxy-galaxy lensing
to cosmic shear is quite sensitive to
the masses of the lenses, the scale at which galaxy-galaxy lensing becomes
unimportant to cosmic shear is relatively independent of the lens masses.
If the results for the galaxy-galaxy lensing contribution to
cosmic shear are extrapolated to large scales, the contribution
of galaxy-galaxy lensing to cosmic shear should vanish for scales
$\theta \gtrsim 5$~arcmin, largely independent of the lens masses.

\section*{Acknowledgments}

I am pleased to thank of Judy Cohen and her collaborators,
especially Mike Pahre and David Hogg, for all of their work on the
Caltech Faint Galaxy Redshift Survey.  Without their efforts, the
work presented in this paper would
not have been possible.  I am also deeply grateful to Judy Cohen for
providing the B-band luminosities of the lens galaxies to me in electronic
form.  In addition, I am very pleased to thank the referee for constructive
comments and helpful suggestions that improved the manuscript.
This work was supported by the National Science Foundation 
under NSF contracts AST-0406844 and AST-0708468.


\clearpage
\begin{thebibliography}{99}

\bibitem[Bartelmann et al. (2001)]{bartelmann et al.} Bartelmann, M. \& Schneider, P. 2001, Phys. Rep., 340, 297

\bibitem[Blandford and Narayan (1986)]{blandford and narayan} Blandford, R. \& Narayan, R. 1986, ApJ, 310, 568

\bibitem[Blandford et al. (1991)]{bsbv} Blandford, R. D., Saust, A.-B., Brainerd, T. G. \& Villumsen, J. V.
1991, MNRAS, 251, 600

\bibitem[Brainerd et al. (1996)]{bbs} Brainerd, T. G., Blandford, R. D. \& Smail, I. 1996, ApJ, 466, 623 (BBS)

\bibitem[Cohen et al. (1996)]{cohen1} Cohen, J. G., Cowie, L. L., Hogg, D. W., Songaila, A., Blandford, R.,
Hu, E. M. \& Shopbell, P. 1996, ApJ, 471, L5

\bibitem[Cohen et al. (2000)]{cohen2} Cohen, J. G., Hogg, D. W., Blandford, R., Cowie, L. L., Hu, E., 
Songaila, A., Shopbell, P., \& Richberg, K. 2000, ApJ, 538, 29

\bibitem[Cohen (2001)]{cohen3} Cohen, J. G. 2001, AJ, 121, 2895

\bibitem[Cooray and Hu (2001)]{cooray and hu} Cooray, A. \& Hu, W. 2001, ApJ, 574, 19

\bibitem[dell'Antonio and Tyson (1996)]{dell'Antonio and Tyson} dell'Antonio, I. P. \& Tyson, J. A. 1996, ApJ, 473, L17

\bibitem[Fischer et al. (2000)]{fischer et al.} Fischer, P. et al.\ 2000, AJ, 120, 1198

\bibitem[Fu et al. (2008)]{fu et al.} Fu, L, Semboloni, E., Hoekstra, H., Kilbinger, M., van Waerbeke, L.,
Tereno, I., Mellier, Y., Heymans, C., Coupon, J., Benabed, K., Benjamin, J., Bertin,
E., Dore, O., Hudson, M. J., Ilbert, O., Maoli, R., Marmo, C., McCracken, H. J. 
\& Menard, B. 2008, AA, 479, 9

\bibitem[Guzik and Seljak (2002)]{guzik and seljak} Guzik, J. \& Seljak, U. 2002, MNRAS, 335, 311

\bibitem[Heymans et al. (2006)]{heymans et al.} Heymans, C., Bell, E. F., Rix, H.-W., Barden, M., Borch, A., Caldwell, J. A. R.,
McIntosh, D. H., Meisenheimer, K., Peng, C. Y., Wolf, C., Beckwith, S. V. W., 
Haussler, B., Jahnke, K., Jogee, S., Sanchez, S. F., Somerville, R. \& Wisotzki, L.
2006, MNRAS, 371, L60

\bibitem[Hilbert et al. (2009)]{hilbert et al.} Hilbert, S., Hartlap, J., White, S. D. M. \& Schneider, P. 2009,
AA, 499, 31

\bibitem[Hoekstra et al. (2004)]{hoekstra1} Hoekstra, H., Yee, H. K. C. \& Gladders, M. D. 2004, ApJ, 606, 67

\bibitem[Hoekstra et al. (2005)]{hoekstra2} Hoekstra, H., Hsieh, B. C., Yee, H. K. C., Lin, H. \& Gladders, M. D.
2005, ApJ, 653, 73

\bibitem[Hogg et al. (2000)]{hogg et al.} Hogg, D. W., Pahre, M. A., Adelberger, K. L., Blandford, R.,
Cohen, J. G., Gautier, T. N., Jarrett, T., Neugebauer, G. \& Steidel, C. C.
2000, ApJS, 127, 1

\bibitem[Howell and Brainerd (2010)]{howell and brainerd} Howell, P. \& Brainerd, T. G. 2010, in preparation

\bibitem[Hudson et al. (1998)]{hudson et al.} Hudson, M. J., Gwyn, S. D. J, Dahle, H. \& Kaiser, N. 1998,
ApJ, 503, 531

\bibitem[Kleinheinrich et al. (2006)]{kleinheinrich et al.} Kleinheinrich, M., Schneider, P., Rix, H.-W., Erben, T., Wolf, C., 
Schirmer, M., Meisenheimer, K., Borch, A., Dye, S., Kovacs, Z. \&
Wisotzki, L. 2006, AA, 455, 441

\bibitem[LeFevre et al. (1996)]{lefevre1} LeFevre, O., Hudon, D., Lilly, S. J., Crampton, D., Hammer, F.
\& Tresse, L. 1996, ApJ, 461, 534

\bibitem[LeFevre et al. (2004)]{lefevre2} LeFevre, O. et al.\ 2004, AA, 428, 1043

\bibitem[Limousin et al. (2007)]{limousin et al.} Limousin, M., Kneib, J.-P., Bardeau, S., Natarajan, P., Czoske, O.,
Smail, I., Ebling, H. \& Smith, G. P. 2007, AA, 461, 881

\bibitem[Lowenthal et al. (1997)]{lowenthal et al.} Lowenthal, J. D., et al.\ 1997 ApJ, 481, 673

\bibitem[Mandelbaum et al. (2006a)]{mandelbaum1} Mandelbaum, R., Seljak, U., Kauffmann, G., Hirata, C. M. \&
Brinkmann, J. 2006a, MNRAS, 368, 715

\bibitem[Mandelbaum et al. (2006b)]{mandelbaum2} Mandelbaum, R., Hirata, C. M., Broderick, T., Seljak, U. \&
Brinkmann, J. 2006b, MNRAS, 370, 1008

\bibitem[Miralda-Escude (1991)]{jordi} Miralda-Escud\'e 1991, ApJ, 370, 1

\bibitem[Munshi et al. (2008)]{munshi et al.} Munshi, D., Valageas, P., van Waerbeke, L. \& Heavens, A. 2008,
Phys. Rep., 462, 67

\bibitem[Natarajan et al. (2009)]{priya} Natarajan, P., Kneib, J.-P., Smail, I., Treu, T., Ellis, R.,
Moran, S., Limousin, M. \& Czoske, O. 2009, ApJ, 693, 970

\bibitem[Parker et al. (2007)]{laura} Parker, L. C., Hoekstra, H., Hudson, M. J., van Waerbeke, L. \&
Mellier, Y. 2007, ApJ, 669, 21

\bibitem[Phillips et al. (1997)]{phillips et al.} Phillips, A. C., Guzman, R., Gallego, J., Koo, D. C., Lowenthal, J. D.,
Vogt, N. P., Faber, S. M. \& Illingworth, G. D. 1997, ApJ, 489, 543

\bibitem[Refregier (2003)]{alex} Refregier, A. 2003, ARAA, 41, 645

\bibitem[Schneider et al. (1992)]{das busch} Schneider, P., Ehlers, J. \& Falco, E. E. 1992, Gravitational
Lenses (Berlin: Springer)

\bibitem[Sheldon et al. (2004)]{sheldon et al.} Sheldon, E., Johnston, D. E., Frieman, J. A., Scranton, R., McKay, T. A.,
Connolly, A. J., Budavari, T., Zehavi, I., Bahcall, N. A., Brinkmann, J. \&
Fukugita, M. 2004, AJ, 127, 2544

\bibitem[Smail et al. (1995)]{smail et al.} Smail, I., Hogg, D. W., Yan, L. \& Cohen, J. G. 1995, ApJ, 449, L105

\bibitem[Steidel et al. (1996)]{steidel et al.} Steidel, C. C., Giavalisco, M., Dickinson, M. \& Adelberger, K. L. 1996,
AJ, 112, 352

\bibitem[Tian et al. (2009)]{tian et al.} Tian, L., Hoekstra, H. \& Zhao, H. 2009, MNRAS, 393, 885

\bibitem[van Waerbeke and Mellier (2003)]{luddo and yannick} van Waerbeke, L. \& Mellier, Y. 2003 (arXive:0305089)

\bibitem[Williams et al. (1996)]{hdf} Williams, R. E., Blacker, B., Dickinson, M., Dixson, W. V. D., 
Ferguson, H. C., Fruchter, A. S., Giavalisco, M., Gilliand, R. L., Heyer, I.,
Katsanis, R., Levay, Z., Lucas, R. A., McElroy, D. B., Petro, L., Postman, M.,
Adorf, H.-M. \& Hook, R. 1996, AJ, 112, 1335

\end{thebibliography}
\end{document}